\documentclass[%
reprint,
superscriptaddress,
 amsmath,amssymb,
 aps,
]{revtex4-2}

\usepackage{cancel}
\usepackage{graphicx}% Include figure files
\usepackage{dcolumn}% Align table columns on decimal point
\usepackage{multirow}
\usepackage{mathtools}
\usepackage[dvipsnames]{xcolor}
\usepackage{booktabs} % For professional table lines
\usepackage{bm}% bold math
\usepackage{braket}
\usepackage[version=4]{mhchem}
\usepackage{float}
\usepackage{hyperref}% add hypertext capabilities
\hypersetup{
    colorlinks=true,      % Set to false if you want colored boxes instead of colored text
    linkcolor=red,       % Color of internal links
    citecolor=blue,       % Color of citation links
    urlcolor=blue,        % Color of external links
    bookmarks=true,       % Show bookmarks bar (required for outline)
    bookmarksopen=true,   % Expand bookmarks tree
    bookmarksnumbered=true % Number bookmarks
}
\begin{document}
\title{Real-Space Quantification of Exciton Localization in Acene Crystals Using Wannier Function Decomposition}% Force line breaks with \\
\author{Zui Tao}
% \altaffiliation{These authors contributed equally to this work.}
\affiliation{Department of Chemistry, University of California Berkeley, Berkeley, California 94720, USA}
% \affiliation{Materials Sciences Division, Lawrence Berkeley National Laboratory, Berkeley, California 94720, USA}
\author{Jonah B. Haber}
% \altaffiliation{These authors contributed equally to this work.}
\affiliation{Department of Materials Science and Engineering, Stanford University, Stanford, CA 94305, USA}
\author{Jeffrey B. Neaton}
\affiliation{Department of Physics, University of California Berkeley, Berkeley, California 94720, USA}
\affiliation{Materials Sciences Division, Lawrence Berkeley National Laboratory, Berkeley, California 94720, USA}
\affiliation{Kavli Energy NanoSciences Institute, Berkeley, California 94720, USA}

\begin{abstract}
We introduce the Wannier function decomposition of excitons (WFDX) method to quantify exciton localization in solids within the \textit{ab initio} Bethe–Salpeter equation framework. By decomposing each Bloch exciton wavefunction into products of single-particle electron and hole maximally localized Wannier functions, this real-space approach provides well-defined orbital- and spatial- resolved measures of both Frenkel and charge-transfer excitons at low computational cost. We apply WFDX to excitons in acene crystals, quantifying how the number of rings, the exciton spin state, and the center-of-mass momentum affect spatial localization. Additionally, we show how this real-space representation reflects structural nonsymmorphic symmetries that are hidden in standard reciprocal-space descriptions. We demonstrate how the WFDX framework can be used to efficiently interpolate exciton expansion coefficients in reciprocal-space and outline how it may facilitate evaluation of observables involving position operators, highlighting its potential as a general tool for both analyzing and computing excitonic  properties in solids.

\end{abstract}

\maketitle

\section{\label{sec:intro}Introduction}
When a material absorbs light, electrons are promoted to higher‐energy states, eliciting a collective response from the remaining electrons. This excited many‐body system can often be mapped onto a bound quasi‐electron and quasi‐hole pair, known as an exciton, stabilized by Coulomb attraction. Quantitative characterization of excitons is essential for understanding absorption spectra, charge-separation efficiency, and other optoelectronic behaviors in many materials. Such excitonic properties underpin the performance of devices ranging from photovoltaic cells and light‐emitting diodes to emerging platforms for optical information storage in transition‐metal dichalcogenides. Excitons are commonly classified by the average spatial separation of their constituent charges: Frenkel excitons \cite{frenkel1931}, with electron and hole confined to the same molecular or atomic site, and charge‐transfer excitons \cite{wannier1937}, in which electron and hole reside on different sites within or even across unit cells. The dominant exciton type has a profound impact on excited-state phenomena such as exciton dynamics \cite{ginsberg2020}: Frenkel exciton transport often proceeds via hopping between sites, whereas charge‐transfer excitons are more likely to diffuse through the crystal lattice like a wave.

First‐principles methods such as time‐dependent density functional theory (TDDFT) \cite{runge1984,yabana1996,petersilka1996,hirata1999,furche2002,onida2002,dreuw2005}, equation‐of‐motion coupled‐cluster (EOM-CC) \cite{stanton1993,nooijen1997a,bartlett2007,krylov2008,zhang2019,gallo2021a,tolle2023,moerman2025a}, and the \textit{ab initio} $GW$–Bethe–Salpeter equation ($GW$–BSE) approach \cite{hybertsen1986,rohlfing2000} now enable quantitative prediction of excitonic properties in both molecular and solid‐state systems. These computational techniques have been successfully applied to molecular crystals \cite{tiago2003,rangel2016,arhangelskis2018,lewis2020,sharifzadeh2012,sharifzadeh2013}, covalent semiconductors \cite{mcclain2017,wing2019,wang2020,moerman2025}, two‐dimensional materials \cite{qiu2013,pulkin2020,camarasa-gomez2023}, and beyond \cite{zhu2014,duan2020,biega2021}. Nevertheless, even with accurate excited‐state solutions in hand, precisely quantifying the extent of electron–hole separation remains challenging, particularly in crystalline solids, where the position operator suffers from an intrinsic gauge ambiguity.

Two widely used approaches to quantify exciton character are natural transition orbitals (NTOs) \cite{martin2003} and exciton correlation functions \cite{sharifzadeh2013}. In the NTO framework familiar from quantum chemistry, one first builds the transition density matrix by projecting the excited state onto diabatic electron and hole orbitals, and then performs a singular-value decomposition to extract the dominant electron–hole pairs. If the leading NTOs localizes electron and hole at opposite ends of a molecule, the exciton is classified as charge-transfer; if they coincide, it is Frenkel-like. This method also readily provides an orbital decomposition for excitons. However, as has been pointed out in Ref. \cite{berkelbach2014}, while NTOs work exceptionally well for molecular or asymmetric systems, they become ambiguous in crystals (or any inversion-symmetric system) because the underlying single-particle orbitals delocalize across the periodic cell, obscuring the difference between the Frenkel and charge-transfer limits. Exciton correlation functions are more suitable for providing real-space picture of electron–hole separation in solids. One fixes the hole at various positions in the unit cell and computes the corresponding average electron density, then visualizes three-dimensional cuts through the resulting six-dimensional wavefunction to reveal spatial separation patterns. This approach offers deep insight but is computationally demanding, requiring sampling of hundreds of hole positions in the unit cell, and a supercell for subsequent analysis of exciton wavefunction. Moreover, although isosurface plots of electron–hole probability are visually informative, they do not identify which specific orbitals contribute most to the exciton as NTOs do.

Here, we introduce a new method, the Wannier function decomposition of excitons (WFDX), for quantifying exciton character. WFDX is computationally cheaper than exciton correlation function approaches and retains orbital (like NTOs) and site resolution for crystalline systems. The idea is motivated by Wannier’s seminal paper \cite{wannier1937} in 1937 where he introduced a localized real-space basis for periodic systems, nowadays called Wannier functions. In addition, he proposed the first application, expanding exciton states in terms of Wannier functions. He showed that one can map correlated electron–hole pairs from reciprocal space into this localized basis, precisely the original motivation for the Wannier functions. In this paper, we build on Wannier’s idea in a modern \textit{ab initio} context. About sixty years after Wannier functions were first written down, the concept was extended to realistic materials with the introduction of maximally-localized Wannier functions (MLWFs) \cite{marzari1997, souza2001}, providing a practical, compact real-space basis for periodic solids \cite{marzari2012, marrazzo2024}. Here, we use the MLWF concept to analyze excitons obtained from \textit{ab initio} many-body perturbation theory, specifically the $GW$ approximation (where $G$ is the one-electron Green’s function and $W$ the screened Coulomb interaction) combined with the Bethe-Salpeter equation (BSE) approach \cite{hedin1965, HedinLundqvist1970, strinati1988, hybertsen1986, rohlfing2000} for excited-state calculations, a standard predictive excited-state framework for solid-state systems. Prior $GW$-BSE studies \cite{deslippe2012} express exciton wavefunctions in a Bloch-function basis, emphasizing their reciprocal-space character. WFDX re-expresses the exciton wavefunction in an electron–hole basis of MLWFs. This real-space formulation naturally captures the character of 
excitons with high orbital and spatial resolution, avoiding both large supercells and an exhaustive sampling of hole positions. 

The remainder of this paper is organized as follows. In Sec.~\ref{sec:xct_k}, we review exciton analysis for solids in reciprocal space using the electron–hole Bloch-function basis. In Sec.~\ref{sec:WFD}, we introduce the WFDX formalism and real-space exciton analysis for solids. Sec.~\ref{sec:impl-com_detail} summarizes the $GW$-BSE methodology and computational details. In Sec.~\ref{sec:result}, we apply WFDX to acene crystals:  we first examine orbital contributions (Sec.~\ref{sec:xct_orbital}), then analyze how the exciton spatial character (Sec.~\ref{sec:xct_spatial}) varies with ring number (Sec.~\ref{sec:num_ring}), spin state (Sec.~\ref{sec:xct_spin}), and center-of-mass momentum $\mathbf{Q}$ (Sec.~\ref{sec:xct_Q}). We further demonstrate how WFDX’s orbital and spatial resolution reflects the presence of nonsymmorphic symmetries (Sec.~\ref{sec:r_ia}). In Sec.~\ref{sec:dis}, we discuss potential future extensions of WFDX, and in Sec.~\ref{sec:conclu}, we conclude.

\section{\label{sec:theory_method}Theory and Methods}

\subsection{\label{sec:xct_k}Exciton Analysis in Reciprocal Space}
Each exciton state $\lvert S\mathbf{Q}\rangle$ in a crystal is uniquely specified by its state index $S$ and center-of-mass (COM) momentum $\mathbf{Q}$.  Its two-particle wavefunction
$\Psi_{S\mathbf{Q}}(\mathbf{r}_e,\mathbf{r}_h)
= \big\langle \mathbf{r}_e,\mathbf{r}_h | S\mathbf{Q} \big\rangle
$
gives the probability amplitude of finding the electron at $\mathbf{r}_e$ and the hole at $\mathbf{r}_h$ simultaneously.

A natural basis for excited-state calculations in periodic systems is the product of single-particle Bloch functions.  Each Bloch state is written as 
$\psi_{n\mathbf{k}}(\mathbf{r})
= \frac{1}{\sqrt{N_{\mathbf{k}}}} e^{i\mathbf{k}\cdot\mathbf{r}}\,u_{n\mathbf{k}}(\mathbf{r}),
$
where $n$ labels the band and $\mathbf{k}$ is the crystal momentum. These one-particle Bloch states can be obtained from Kohn–Sham DFT, Hartree–Fock, or any other electronic structure method suitably formulated for periodic systems. In this basis the exciton wavefunction is expanded as
\begin{equation}
\Psi_{S\mathbf{Q}}(\mathbf{r}_e,\mathbf{r}_h)
= \sum_{c v \mathbf{k}}
A_{cv\mathbf{k}}^{S\mathbf{Q}}\,
\psi_{c\mathbf{k}}(\mathbf{r}_e)\,
\psi_{v,\mathbf{k}-\mathbf{Q}}^{\star}(\mathbf{r}_h),
\label{eq:xct_k}
\end{equation}
where $c$ and $v$ index conduction (virtual) and valence 
 (occupied) bands, respectively, and $A_{cv\mathbf{k}}^{S\mathbf{Q}}$ are the exciton expansion coefficients.

\begin{figure*}[!htb]
  \centering
  \includegraphics[width=0.7\linewidth]{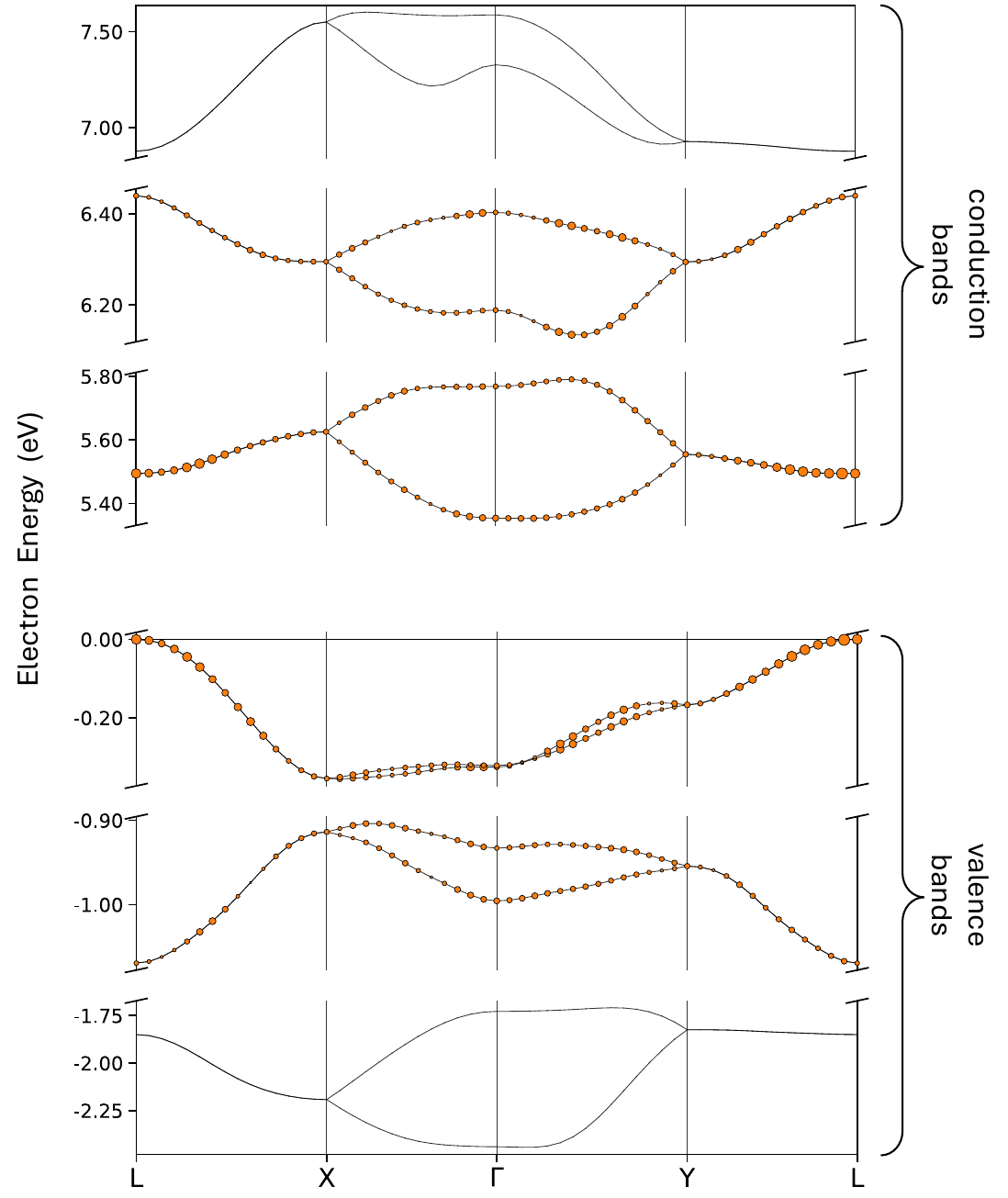}
  \caption{\label{fig:xct_k}
    Reciprocal-space decomposition of the lowest-energy zero-COM-momentum singlet exciton in crystalline naphthalene.  The size of the data point at each $\mathbf{k}$ is a reflection of the state's contribution: for a valence state $\left|v\mathbf{k} \right \rangle$, size $\propto \sum_{c}|A_{cv\mathbf{k}}^{S\mathbf{Q}}|^2$; for a conduction state $\left|c\mathbf{k} \right \rangle$, size $\propto \sum_{v}|A_{cv\mathbf{k}}^{S\mathbf{Q}}|^2$. The exciton expansion coefficients are computed on a uniform \textbf{k}-gird and interpolated onto the selected high-symmetry \textbf{k}-path via Eq.~\ref{eq:iFFT}, see Sec.\ref{sec:dis} for details. The underlying electronic band structure shows $GW$ quasiparticle energies referenced to the valence band maximum (VBM = $0$ eV). For clarity, each pair of bands is displayed in a vertically zoomed window with its own energy scale; the y-axis is broken between windows.}
\end{figure*}

As shown in Fig.~\ref{fig:xct_k}, one can analyze exciton character in reciprocal space by examining the magnitudes of $A_{cv\mathbf{k}}^{S\mathbf{Q}}$, which indicate how the Bloch states $\left | n \mathbf{k} \right \rangle$ contribute to a given exciton. However, because Bloch states extend over the entire crystal, this analysis does not yield a direct real-space picture of electron–hole separation.  In the next section, we introduce a localized basis transformation that enables quantitative analysis of excitons in real space.

\subsection{\label{sec:WFD}Wannier Function Decomposition of Excitons}

The extended single particle Bloch basis $\psi_{n\mathbf{k}}(\mathbf{r})=\left\langle\mathbf{r}|n\mathbf{k}\right\rangle$ and the localized Wannier basis \(w_{m\bar{\mathbf{R}}}(\mathbf{r})=\langle \mathbf{r}| m\bar{\mathbf{R}}\rangle\) with orbital index $m$ are related through a unitary transformation, via 
\begin{subequations}
\begin{align}
w_{m \bar{\mathbf{R}}}(\mathbf{r})
  &= \frac{1}{\sqrt{N_{\mathbf{k}}}}
     \sum_{n \mathbf{k}}^{N_{\mathcal{W}}}
     e^{-i \mathbf{k}\cdot \bar{\mathbf{R}}}
     U_{nm}(\mathbf{k}) \psi_{n\mathbf{k}}(\mathbf{r})
     \label{eq:WF-Bloch} \\
\shortintertext{and}
\psi_{n \mathbf{k}}(\mathbf{r})
  &= \frac{1}{\sqrt{N_\mathbf{k}}}
     \sum_{m \bar{\mathbf{R}}}^{N_{\mathcal{W}}}
     e^{i \mathbf{k}\cdot \bar{\mathbf{R}}}
     U_{mn}^\dagger(\mathbf{k})\, w_{m \bar{\mathbf{R}}}(\mathbf{r}).
     \label{eq:Bloch-WF}
\end{align}
\end{subequations}
Here \(\bar{\mathbf{R}}\) is a lattice vector and labels the cell where Wannier function is centered, \(N_{\mathbf{k}}\) is the total number of \(\mathbf{k}\)-points, and \(\mathcal{W}\) denotes the chosen subspace of bands (dimension \(N_{\mathcal{W}}\)).  The unitary matrices \(U(\mathbf{k})\) encode a gauge freedom and in the MLWF framework, this gauge is specified by minimizing the total spread functional  
\begin{equation}
\Omega[U]=\sum_m^{N_{\mathcal{W}}}\left[\langle m \bar{\mathbf{0}}| \hat{r}^2|m \bar{\mathbf{0}}\rangle-\langle m \bar{\mathbf{0}}| \hat{\mathbf{r}}|m \bar{\mathbf{0}}\rangle^2\right],
\label{eq:FB}
\end{equation}
following the procedure implemented in the open source, post-processing software \texttt{Wannier90} \cite{mostofi2014, pizzi2020a}. In Eq.~\ref{eq:FB},  $\hat{\mathbf{r}}$ is understood as the position operator with three components and the notation $\Omega[U]$ indicates that the spread is a functional of the gauge, $U$. 

Substituting Eq.~\ref{eq:Bloch-WF} into the reciprocal-space expression in Eq.~\ref{eq:xct_k} and collecting terms yields the exciton wavefunction in the Wannier basis (see Appendix~\ref{sec:coefficient_k2R}), namely
\begin{equation}
\begin{aligned}
    &\Psi_{S \mathbf{Q}}\left(\mathbf{r}_e, \mathbf{r}_h\right) \\
    &= \frac{1}{\sqrt{N_\mathbf{k}}}\sum_{\bar{\mathbf{R}}} e^{i \mathbf{Q} \cdot \bar{\mathbf{R}}}
     \left[ \sum_{a i \Delta \bar{\mathbf{R}}} A_{a i \Delta \bar{\mathbf{R}}}^{S \mathbf{Q}} w_{a \bar{\mathbf{R}}+\Delta \bar{\mathbf{R}}}\left(\mathbf{r}_e\right) w_{i \bar{\mathbf{R}}}^{\star}\left(\mathbf{r}_h\right)\right],
    \label{eq:xct_wannier}    
\end{aligned}    
\end{equation}
where the indices \(i\) and \(a\) label occupied (valence) and virtual (conduction) Wannier orbitals, respectively, and \(\Delta\bar{\mathbf{R}}\) is the lattice-vector difference between the cell where the electron and hole MLWFs are centered.  We define the electron–hole Wannier center difference, which measures the separation between the MLWF centers of the electron and the hole, as
\begin{equation}
    \Delta \mathbf{R}=\Delta\bar{\mathbf{R}}+\boldsymbol{\tau}_a-\boldsymbol{\tau}_i, 
\end{equation}
with $\boldsymbol{\tau}_a$ and $\boldsymbol{\tau}_i$ the Wannier centers of the virtual (unoccupied) and occupied orbital in the home cell labelled by \(\bar{\mathbf{R}}=\mathbf{0}\).  Analogous to reciprocal-space “dot plots” shown in Fig. \ref{fig:xct_k}, one can visualize WFDX with arrow plots (Fig.~\ref{fig:A_B_naph_arrow}), where each arrow originates at the hole’s Wannier center and points to the electron’s center.  $\Delta \mathbf{R}$ is the separation between the electron-hole Wannier center. In analogy to the reciprocal space exciton analysis, $|A_{a i \Delta \bar{\mathbf{R}}}^{S \mathbf{Q}}|^2$ is the probability of having the transition from occupied orbital $i$ to the virtual orbital $a$ with an electron-hole spatial separation of $\Delta \mathbf{R}$ for exciton state $|S\mathbf{Q}\rangle$. This provides a corresponding real space analysis for excitons in periodic solids, and we coin this as the Wannier function decomposition of excitons (WFDX) approach. 

\section{\label{sec:impl-com_detail}Implementation and Computational Details}
The WFDX method can be applied to analyze any exciton wavefunction within Tamm-Dancoff approximation (TDA) \cite{dancoff1950,onida2002},  standard within the \textit{ab initio} $GW$–BSE framework \cite{rohlfing2000}; the key steps in the WFDX implementation are summarized below.

\subsection{\label{sec:impl}Implementation}
Exciton coefficients $A_{c v \mathbf{k}}^{S\mathbf{Q}}$ are first obtained by solving the BSE in reciprocal space within the TDA: 
\begin{equation}
\begin{aligned}
    &\left(\varepsilon_{c \mathbf{k}}^{GW}-\varepsilon_{v \mathbf{k}-\mathbf{Q}}^{GW}\right) A_{c v \mathbf{k}}^{S \mathbf{Q}} \\
 &+ \sum_{c^{\prime} v^{\prime} \mathbf{k}^{\prime}}\langle c v \mathbf{k} \mathbf{Q}| K^{\mathrm{eh}}\left|c^{\prime} v^{\prime} \mathbf{k}^{\prime} \mathbf{Q}\right\rangle A_{c^{\prime} v^{\prime} \mathbf{k}^{\prime}}^{S \mathbf{Q}}=E_{S \mathbf{Q}} A_{c v \mathbf{k}}^{S \mathbf{Q}},
 \end{aligned}
\end{equation}
where $\varepsilon_{n \mathbf{k}}^{GW}$ are the $GW$ quasiparticle energies and $K^{\mathrm{eh}}=-K^{\mathrm{D}}+2\delta_SK^{\mathrm{X}}$ is the electron-hole interaction kernel. $K^{\textrm{D}}$ is the direct kernel capturing the screened attraction between electron and hole; $K^{\mathrm{X}}$ is the exchange kernel in analogous to the Hartree-Fock exchange interaction, and it only exists for singlet excitons with zero exciton spin. $E_{S\mathbf{Q}}$ is the exciton energy.

In the WFDX method, the real-space exciton coefficients
$
A_{a i\,\Delta\bar{\mathbf{R}}}^{S\mathbf{Q}}
$
are obtained from the reciprocal-space amplitudes via
% \begin{widetext}
\begin{equation}
        A_{a i \Delta \bar{\mathbf{R}}}^{S \mathbf{Q}} =\frac{1}{\sqrt{N_{\mathbf{k}}}} \sum^{N_{\mathcal{W}}}_{cv\mathbf{k}} e^{i \mathbf{k} \cdot \Delta \bar{\mathbf{R}}} U_{a c}^{\dagger}(\mathbf{k}) \, A_{c v \mathbf{k}}^{S \mathbf{Q}} \, U_{v i}(\mathbf{k}-\mathbf{Q}).
    \label{eq:A_coeff_wannier}
\end{equation}
% \end{widetext}
% \begin{equation}
%         \nabla_{\mathbf{Q}} A_{cv\mathbf{k}}^{S\mathbf{Q}}=\frac{1}{N_{\mathbf{Q}}}\sum_{M\mathbf{R} S' \mathbf{Q}'} e^{i(\mathbf{Q}-\mathbf{Q}')\cdot \mathbf{R}} A_{cv\mathbf{k}}^{S'\mathbf{Q}'} \left[ i\mathbf{R} \, V^{\dagger}_{MS}(\mathbf{Q})V_{S'M}(\mathbf{Q}') \, + V_{S'M}(\mathbf{Q}') \, \nabla_{\mathbf{Q}} V^{\dagger}_{MS}(\mathbf{Q})\right] 
% \end{equation}
% \begin{equation}
%     A_{cv\mathbf{k}}^{S\mathbf{Q}}=\frac{1}{N_{\mathbf{Q}}}\sum_{M\mathbf{R} S' \mathbf{Q}'} e^{i(\mathbf{Q}-\mathbf{Q}')\cdot \mathbf{R}} V^{\dagger}_{MS}(\mathbf{Q})A_{cv\mathbf{k}}^{S'\mathbf{Q}'}V_{S'M}(\mathbf{Q}')
% \end{equation}
Here $U_{ac}(\mathbf{k})$ and $U_{vi}(\mathbf{k}-\mathbf{Q})$ are the unitary rotations that generate MLWFs for conduction (electron or virtual) and valence (hole) subspaces at each $\mathbf{k}$-point, computed via the standard single-particle wannierization in \texttt{Wannier90} \cite{mostofi2014, pizzi2020a}. For clarity we display the conduction and valence rotations separately above; in practice a single unitary $U(\mathbf{k})$ rotates the entire composite manifold (valence + conduction) into the Wannier gauge, so the electron and hole subspaces need not be treated separately. The real-space exciton coefficients can then be written in the compact form
\begin{equation}
    A_{mm' \Delta \bar{\mathbf{R}}}^{S \mathbf{Q}} = \frac{1}{\sqrt{N_{\mathbf{k}}}} \sum_{nn'\mathbf{k}}^{N_{\mathcal{W}}} e^{i\mathbf{k} \cdot \Delta \bar{\mathbf{R}}} \, U^{\dagger}_{mn}(\mathbf{k}) \, A_{nn'\mathbf{k}}^{S\mathbf{Q}} \, U_{n'm'}(\mathbf{k-Q}),\label{eq:A_r_nm}
\end{equation}
where $m, m'$ label Wannier orbitals, and $n, n'$ label Bloch bands in the chosen composite subspace. $A_{nn'\mathbf{k}}^{S\mathbf{Q}}$ is the zero-padded CV-block embedding of the usual TDA amplitudes: $A_{nn'\mathbf{k}}^{S\mathbf{Q}}=A_{cv\mathbf{k}}^{S\mathbf{Q}}$ for $n \in C \text{(conducation)}$, $n' \in V \text{(valence)}$, and zero otherwise.

% in practice, they are combined into the composite transformation shown in Eq.~\ref{eq:A_r_nm} (Appendix~\ref{sec:coefficient_k2R}).

\begin{figure}[!htb]
\includegraphics[width=1\linewidth]{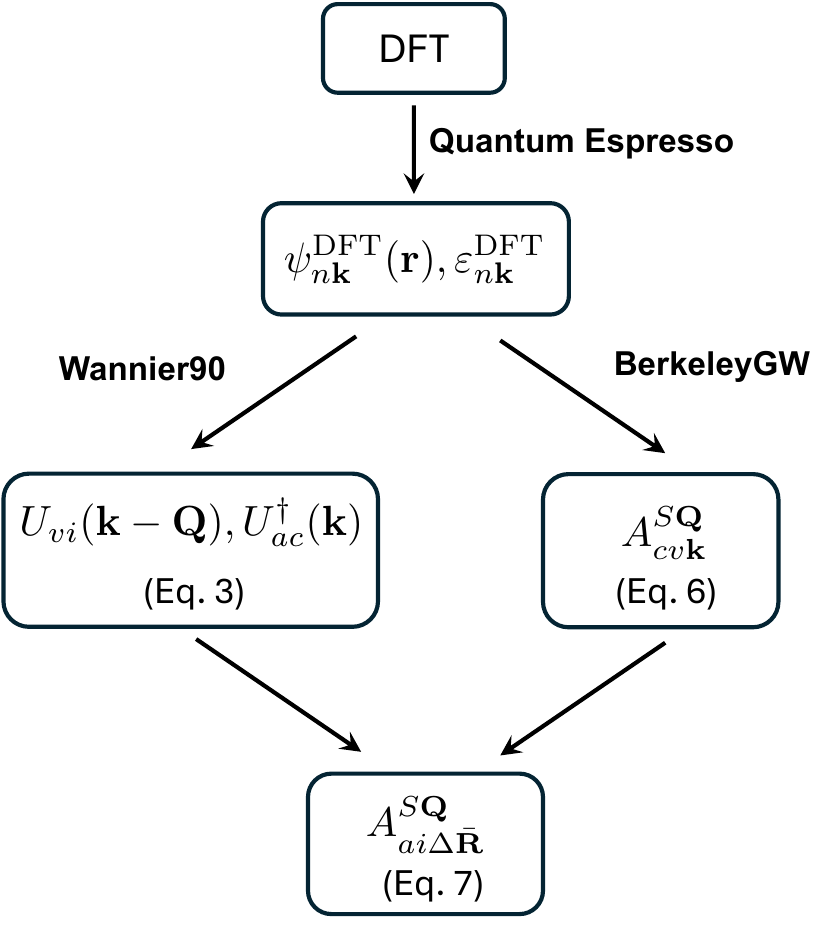}
\caption{\label{fig:workflow}Workflow for WFDX in the $G_{0}W_{0}$–BSE approach.  Left: wannierization of valence and conduction subspaces to obtain $U(\mathbf{k})$.  Right: BSE solution for $A_{cv\mathbf{k}}^{S\mathbf{Q}}$. Software used in this work is indicated, but the workflow is independent of codes.}
\end{figure}

Fig.~\ref{fig:workflow} outlines the overall workflow. For the exciton states valid under the one-shot $GW$ approximation ($G_0W_0$), the quasiparticle energies $\varepsilon_{n \mathbf{k}}^{GW}$ are corrected to first order in perturbation theory and the quasiparticle wavefunctions remain at zeroth order. One can start the wannierization from the DFT eigenstates in this case as shown in the workflows. This scheme can be 
 readily extended beyond one-shot $GW$; for instance, one may use quasiparticle energies and wavefunctions from an eigenvalue or eigenvector self-consistent $GW$ scheme \cite{vanschilfgaarde2006, bruneval2006}, incorporating higher-order corrections. These same quasiparticle states can be used directly as input for wannierization and be employed as an electron–hole product basis for expanding BSE-computed exciton wavefunctions.

\subsection{\label{sec:com_detail}Computational Details}
In what follows, we use WFDX to study naphthalene through pentacene (2–5 rings) crystals using experimental structures from the Cambridge Structural Database: NAPHTA31 \cite{capelli2006}, ANTCEN16 \cite{chaplot1982}, TETCEN03 \cite{robertson1961}, and PENCEN \cite{campbell1962}. Acenes are linearly fused benzene rings, and in solid state structures we study, they all adopt a herringbone packing arrangement. Each acene crystal unit cell contains two molecules oriented differently, one at the corner (site A) and one at the face center (site B) as shown in Fig. \ref{fig:naph_band_orbital}. All four acene crystals studied possess both identity and inversion symmetries. Naphthalene and anthracene assume monoclinic lattices, introducing two more glide mirror symmetries that interchange the A and B sublattices; tetracene and pentacene, by contrast, crystallize in triclinic cells lacking these operations, rendering their two molecular sites inequivalent. We will show how our WFDX analysis reflects those additional nonsymmorphic symmetries in Sec. \ref{sec:r_ia}.

The starting point for all calculations is a ground-state density functional theory (DFT) calculation to obtain Kohn-Sham energies and eigenstates \cite{hohenberg1964, kohn1965}. We use a plane wave basis set, the Perdew-Burke-Ernzerhof (PBE) exchange-correlation functional \cite{perdew1997}, and norm-conserving pseudopotentials taken from \texttt{pseudo-dojo} \cite{hamann2013, vansetten2018}. To converge the ground-state density we use an 70 Ry planewave cutoff and $8\times8\times4$ \textbf{k}-mesh. All DFT calculations are performed with \texttt{Quantum Espresso} \cite{giannozzi2017}.  

On top of these Kohn–Sham states we perform one-shot $G_{0}W_{0}$ and subsequently solve the BSE. Based on prior benchmarks \cite{rangel2016}, for acenes up to five rings, $G_{0}W_{0}$ with a PBE starting point yields reliable quasiparticle energies. In the random phase approximation (RPA) for the dielectric matrix \cite{adler1962, wiser1963}, we include a total of 400 bands and use an 8 Ry planewave cutoff to compute the susceptibility. Frequency dependence is treated with the Hybertsen–Louie generalized plasmon‐pole (GPP) model \cite{hybertsen1986}. The BSE kernel, $K^{\mathrm{eh}}$, employs static screening in the direct term and is expanded over 16 conduction bands and 16 valence bands. We solve the exciton COM momentum ($\mathbf{Q}$) dependent BSE, constructing the electron-hole kernel explicitly along selected \textbf{Q}-path without symmetry reduction or interpolation. All $G_0W_0$-BSE steps are carried out with \texttt{BerkeleyGW} \cite{deslippe2012}. 

Details on the construction of the MLWFs are provided in Appendix~\ref{sec:acene_band_MLWF}.

\section{\label{sec:result}Results}
As a first demonstration, we apply WFDX to quantify exciton localization in herringbone acene crystals \cite{tiago2003, sharifzadeh2013, rangel2016}, a class of organic semiconductors \cite{sharifzadeh2012} of great interest where previous work have reported on the carrier mobility \cite{lee2018, brown-altvater2020}, exciton transport \cite{muller2023, cohen2024}, singlet fission \cite{berkelbach2014, refaely-abramson2017, altman2022} and other emerging phenomena \cite{brown-altvater2016, zhang2023, alvertis2023}. Accurate real-space characterization of exciton spatial extent and orbital contributions in these systems is essential for understanding their optoelectronic behavior.

\subsection{\label{sec:xct_orbital}Exciton Orbital Character}
\begin{figure*}[!htb]
\includegraphics[width=0.8\linewidth]{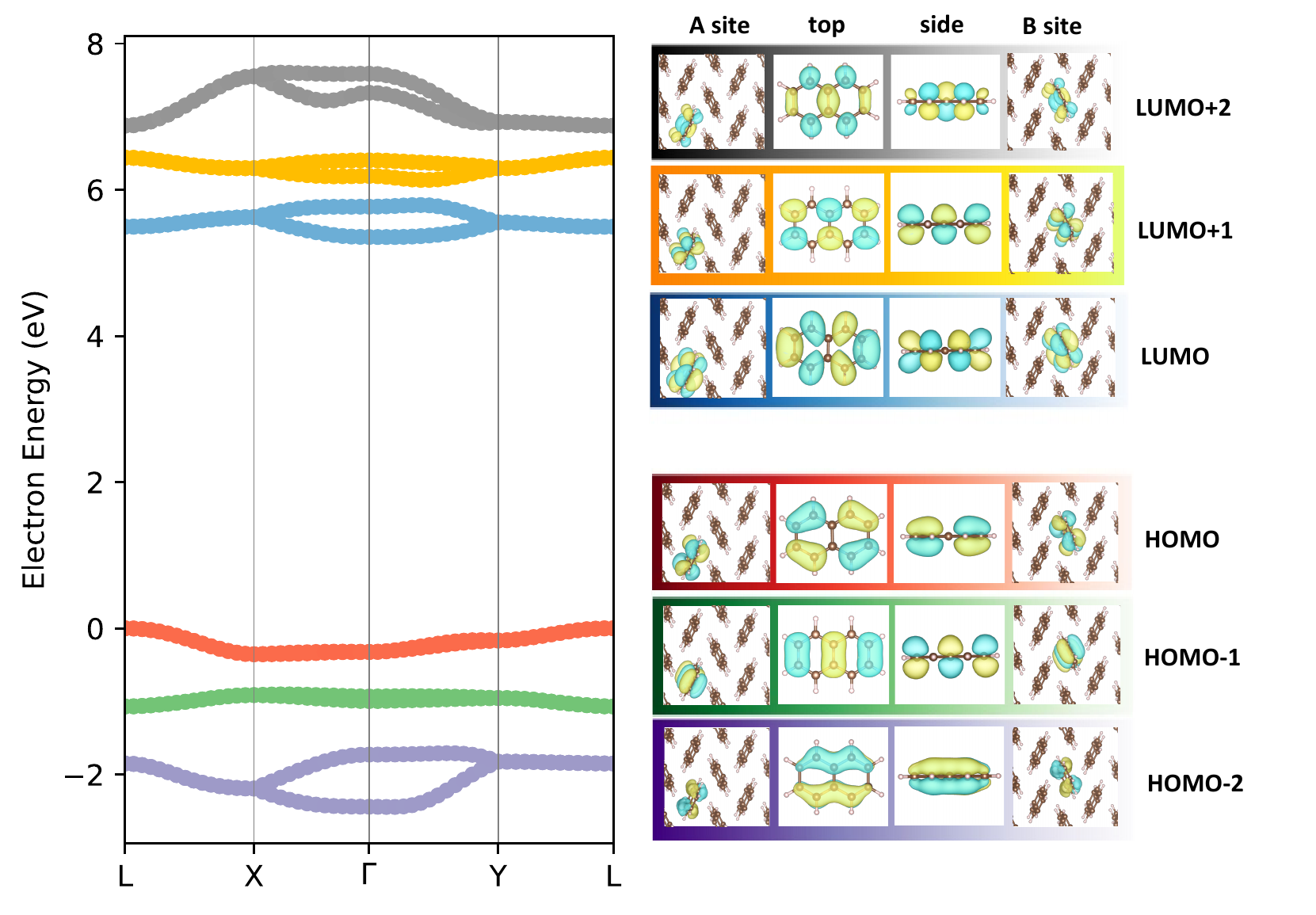}
\caption{\label{fig:naph_band_orbital} Left: $GW$ quasiparticle band structure of crystalline naphthalene, referenced to the valence band maximum (VBM energy set to 0 eV). Curves are colored by MLWF sublattice character (dark: site A; light: site B).  The near-midpoint colors across the bands indicate that each Bloch state is approximately an equal-weight superposition of MLWFs on the A and B sublattices.
Right: Isosurfaces of the MLWFs for HOMO–2, HOMO–1, HOMO, LUMO, LUMO+1 and LUMO+2, columns from left to right show A site, top, side and B site views.}
\end{figure*}

In what follows, we use the fact that \(A_{a i \Delta\mathbf{R}}^{S\mathbf{Q}}\equiv A_{a i \Delta\bar{\mathbf{R}}}^{S\mathbf{Q}}\) and express all amplitudes in terms of $\Delta \mathbf{R}$, since once $i$, $a$ and $\Delta \bar{\mathbf{R}}$ are specified, $\Delta \mathbf{R}$ is uniquely determined and the probability amplitude interpretation is unchanged. We retain the indices $a$ and $i$ to specify the exact electron-hole transition and its localization site. This notation preserves both clarity and conciseness. The orbital contribution to an exciton from a transition between occupied Wannier orbital \(i\) and virtual Wannier orbital \(a\) is written as

\begin{equation}
    \mathcal{P}_{i\rightarrow a}^{S \mathbf{Q}}=\sum_{\Delta \mathbf{R}}\left|A_{a i \Delta \mathbf{R}}^{S \mathbf{Q}}\right|^2.
    \label{eq:P_ia}
\end{equation}
Fig.~\ref{fig:naph_band_orbital} illustrates the band structure of crystalline naphthalene, color-coded by MLWF character alongside the isosurfaces of the MLWFs. Similar plots for the other acenes are provided in Appendix~\ref{sec:acene_band_MLWF}.

As mentioned in Sec.~\ref{sec:com_detail}, because the acene unit cell contains two inequivalent molecules (sites A and B), each molecular orbital generates a pair of crossing bands and corresponding Wannier functions on the two molecular sublattices. We then group the site-specific Wannier functions into molecular-orbital labels, denoted \(I\) for occupied groups (e.g. A HOMO and B HOMO are grouped as HOMO) and \(A\) for virtual groups (e.g. A LUMO and B LUMO are grouped as LUMO), so that
\begin{equation}
    \mathcal{P}_{I\rightarrow A}^{S \mathbf{Q}}=\sum_{i \in I,\, a \in A}\mathcal{P}_{i\rightarrow a}^{S \mathbf{Q}}.
    \label{eq:P_IA}
\end{equation}

For the lowest singlet exciton at \(\mathbf{Q}=\mathbf{0}\), the dominant transitions in solid naphthalene are HOMO\(-1\)\(\to\)LUMO and HOMO\(\to\)LUMO\(+1\), rather than the simple HOMO\(\to\)LUMO (Table~\ref{tab:naph_orbital}). This is consistent with a prior excited state orbital transition study of the gas-phase naphthalene molecule in Ref. \cite{kimber2023}. In contrast, solid anthracene, tetracene, and pentacene exhibit nearly 100\% HOMO\(\to\)LUMO character for singlets, and all triplet excitons in these acene solids are essentially dominated by HOMO\(\to\)LUMO transitions.  
\begin{table}[!htb]
\caption{Orbital contributions to the lowest singlet exciton (\(\mathbf{Q}=\mathbf{0}\)) in acene crystals.}
\begin{ruledtabular}
\begin{tabular}{ccc}
Number of rings & Transition              & Contribution (\%)  \\ \midrule
\multirow{4}{*}{2}              & HOMO-1 $\to$ LUMO            & 37                     \\ 
                                % & HOMO-1 $\to$ LUMO+1          & 2                      \\
                                & HOMO $\to$ LUMO              & 23                     \\ 
                                & HOMO $\to$ LUMO+1            & 38                     \\ 
                                & Total       & $\simeq$100 \\ \hline
3-5                             & HOMO $\to$ LUMO              & $\simeq$100            \\              
\end{tabular}
\end{ruledtabular}
\label{tab:naph_orbital}
\end{table}

\subsection{\label{sec:xct_spatial}Exciton Real-Space Character}
The probability of finding the electron–hole pair separated by \(\Delta\mathbf{R}\) in exciton \(\lvert S\mathbf{Q}\rangle\) is
\begin{equation}
    \mathcal{P}_{\Delta \mathbf{R}}^{S \mathbf{Q}}=\sum_{ai}\left|A_{a i \Delta \mathbf{R}}^{S \mathbf{Q}}\right|^2.
    \label{eq:P_R}
\end{equation}
We define the Frenkel character as the on‐site weight at zero separation,
$\mathcal{P}_{\text{F}}^{S \mathbf{Q}}=\mathcal{P}_{\Delta \mathbf{R}=\mathbf{0}}^{S \mathbf{Q}};
    \label{eq:P_FR}
$
and the charge-transfer character as the sum over all nonzero separations,
$
    \mathcal{P}_{\text{CT}}^{S \mathbf{Q}}=\sum_{\Delta \mathbf{R} \neq \mathbf{0}}\mathcal{P}_{\Delta \mathbf{R}}^{S \mathbf{Q}}
    \label{eq:P_CT}
$. This decomposition offers a clear cut distinction between Frenkel and charge-transfer excitons. In contrast, NTOs require examining the spatial overlap of electron and hole orbitals, while exciton correlation functions rely on an extra real‐space integration to extract charge‐transfer weight, introducing a degree of arbitrariness to the metric. In WFDX, once the electron and hole MLWFs are determined, the exciton wavefunction admits a unique expansion in that localized basis, and the Frenkel versus charge‐transfer contributions follow directly from the electron–hole Wannier centers.

To quantify the exciton’s spatial extent, we compute the root‐mean‐square (RMS) electron–hole separation,
\begin{equation}
    \sqrt{\braket{|\Delta\mathbf{R}|^2}}=\sqrt{\sum_{\Delta\mathbf{R}} \mathcal{P}_{\Delta \mathbf{R}}^{S\mathbf{Q}} \, |\Delta \mathbf{R}|^2}.
    \label{eq:eh-distance}
\end{equation}
WFDX is not limited to acenes but offers a general real‑space framework for quantifying exciton character in any crystalline solid. In the Wannier–Mott limit of a hydrogenic exciton, the RMS separation of Eq.~\ref{eq:eh-distance} would correspond physically to a Bohr radius; yet the same definition applies equally to non‑hydrogenic excitons, making it a universally valid measure of exciton size.

\subsubsection{\label{sec:num_ring}Dependence on Acene Length}
\begin{figure*}[!htb]
\includegraphics[width=0.9\linewidth]{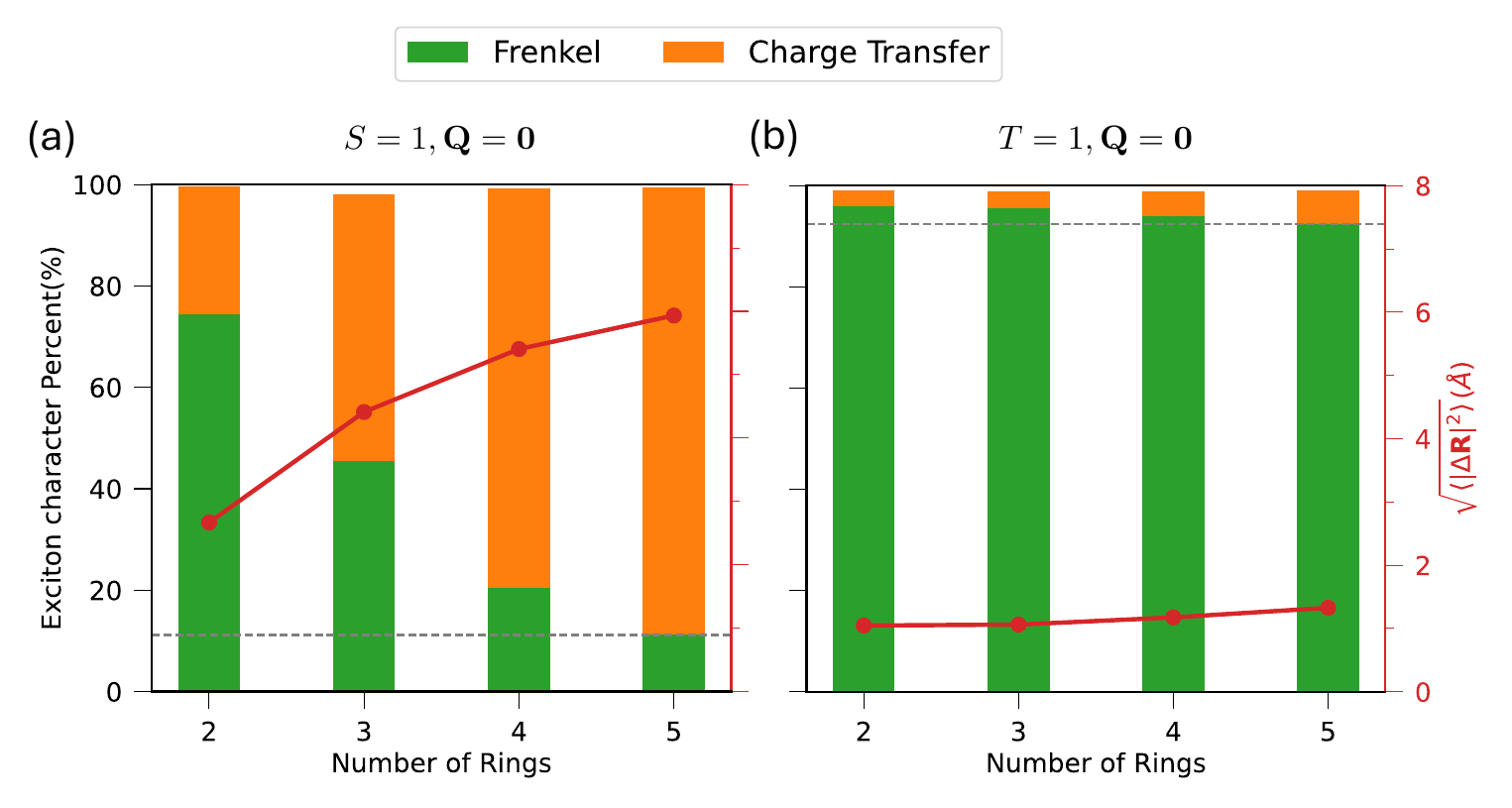}
\caption{\label{fig:acene_S-T} Trends in lowest-lying singlet and triplet excitons in acene solids. Frenkel (\(\mathcal{P}_\mathrm{F}\), green bars) versus charge-transfer (\(\mathcal{P}_\mathrm{CT}\), orange bars) character (left black axis) and root-mean-square electron–hole separation (red points and right red axis) for the lowest-energy \(\mathbf{Q}=\mathbf{0}\) (a) singlet and (b) triplet excitons, plotted against the number of rings in the acene molecule making up the crystal.}
\end{figure*}

Figure~\ref{fig:acene_S-T} shows the Frenkel and charge‐transfer percentages (\(\mathcal{P}_{\mathrm F}\) and \(\mathcal{P}_{\mathrm{CT}}\)) on the left axis and the RMS separation on the right axis for the lowest‐energy singlet and triplet excitons. We find that as the number of rings increases from naphthalene to pentacene, the exciton becomes progressively more delocalized: both the charge-transfer character and the RMS electron–hole distance grow as physically expected. This is consistent with the larger electronic bandwidth and therefore intermolecular coupling as the molecules increase in number of rings, see Appendix~\ref{sec:acene_band_MLWF}, \ref{sec:acene_coupl}.  In pentacene, the singlet exciton’s average separation is approximately 6 \AA, in excellent agreement with the 6.1 \AA \,peak reported via correlation functions in Ref.~\cite{sharifzadeh2013} (the small discrepancy reflects different quantification schemes).

\subsubsection{\label{sec:xct_spin}Singlet vs.\ Triplet Excitons}
We denote singlet states by $S$ and triplets by $T$.  In Fig.~\ref{fig:acene_S-T}(a), the lowest singlet in naphthalene is predominantly Frenkel-like, but from anthracene onward it becomes charge-transfer dominated, with the charge-transfer fraction increasing and reaching $\sim 10\%$ in pentacene.  By contrast (Fig.~\ref{fig:acene_S-T}(b)), all lowest triplet excitons remain strongly Frenkel-like, with only a slight increase in charge-transfer weight as the ring count grows.  From these results we can infer that in the acene crystals exciton delocalization is driven primarily by the repulsive exchange interaction and intermolecular coupling.

\subsubsection{\label{sec:xct_Q} $\mathbf{Q}$-Dependence of Exciton Character}
By solving the BSE for different exciton COM momentum $\mathbf{Q}$, we calculate the exciton dispersion along the x direction at 5 equally spaced $\mathbf{Q}$ points from $\Gamma$ to $X$. We focus on anthracene as it is representative for studying the exciton spatial character change with respect to $\mathbf{Q}$: it is the system with the least number of rings for which the lowest four exciton bands start to cross, and it preserves the nonsymmorphic symmetry, thereby accentuating the monotonic evolution of exciton character along the bands connected by maximizing the overlap of eigenstates at adjacent $\mathbf{Q}$-points. The analysis for other acene crystals are given in Appendix~\ref{sec:Q_xct_dispersion}.

As shown in Fig.~\ref{fig:anth_Q}, the lowest four exciton bands of crystalline anthracene are entangled but separated from higher energy states. States $S=1,3$ and $S=2,4$ are degenerate on the $X$ point due to the nonsymmorphic symmetry. Overall, all of the lowest-lying anthracene excitons are charge-transfer-dominated, and $S=1,3$ have relatively more Frenkel character compared with $S=2,4$. For $S=1$, the average electron-hole distance increases modestly with increasing charge-transfer character, and its counterpart $S=3$ is the opposite. 

Notably, exciton band dispersion does not directly correlate with real-space localization. We attribute this to the fast that our WFDX analyzes only the relative coordinate $\Delta\mathbf{R}$ via single-particle Wannier functions. By contrast the more dispersive exciton bands are predicted to correlate with localization of the exciton in the COM coordinate, as defined in Ref. \cite{haber2023}.

\begin{figure*}[!htb]
\includegraphics[width=1\linewidth]{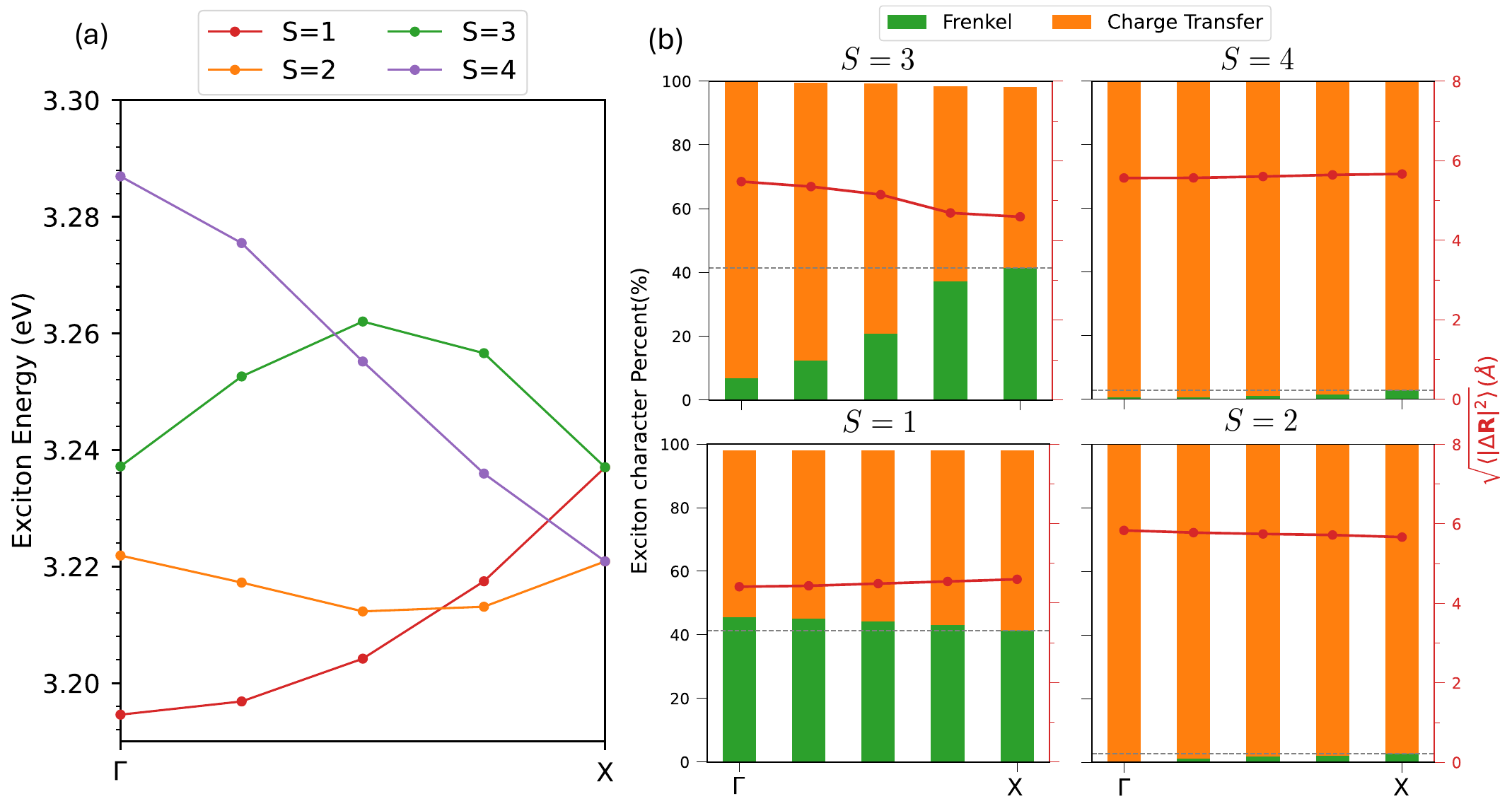}
\caption{\label{fig:anth_Q} (a) Exciton dispersion for singlet in crystal anthracene. Connectivity of the bands are chosen to maximize the overlap of eigenstates at adjacent \textbf{Q}-points. (b) The spatial character change for the lowest four exciton along the \textbf{Q}-path.}
\end{figure*}

\subsection{\label{sec:r_ia}Orbital and Spatial Resolution}
To combine orbital and spatial information, we introduce arrow diagrams to plot the real‐space amplitudes \(A_{a i \Delta \mathbf{R}}^{S\mathbf{Q}}\) from Eq.~\ref{eq:xct_wannier}.  Because the \(c\) axis is longer than $a$ or $b$ in these acenes, nearly all electron–hole separations are confined to the same \(ab\) plane, so we project each arrow plot onto that plane. 

\begin{figure*}[!htb]
\includegraphics[width=1\linewidth]{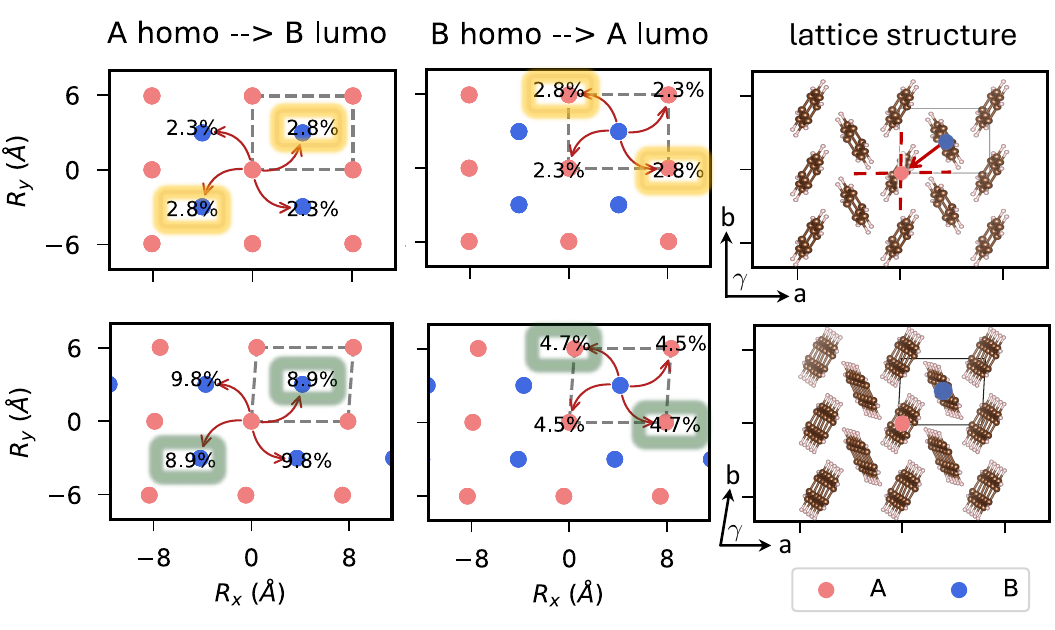}
\caption{\label{fig:A_B_naph_arrow}
HOMO\(\to\)LUMO transition probabilities for the lowest‐energy singlet exciton at \(\mathbf{Q}=\mathbf{0}\) in crystalline naphthalene (top row) and pentacene (bottom row).  Columns~1 and~2 show transitions originating from sublattice A\(\to\)B and B\(\to\)A, respectively; pink dots indicate Wannier centers on sublattice A and blue dots those on sublattice B. For clarity, only nearest‐neighbor contributions are shown; full‐range plots appear in Appendix~\ref{sec:acene_arrow_plot}. In naphthalene, yellow boxes connect equal probabilities related by the glide‐mirror symmetry, whereas in pentacene green boxes highlight the inequivalent values that arise when that symmetry is absent.  The rightmost column gives each crystal's \(ab\)-plane projection, with the arrow indicating the glide translation \(\tau\) and dashed lines denoting the mirror planes \(\sigma\) that together generate the glide‐mirror operations.
}
\end{figure*}

Fig.~\ref{fig:A_B_naph_arrow} shows A \(\to\) B and B \(\to\) A HOMO \(\to\) LUMO transitions for the lowest-energy exciton at \(\mathbf{Q}=\mathbf{0}\) in crystal naphthalene and pentacene; results for other transitons and acenes are in Appendix~\ref{sec:acene_arrow_plot}. Each arrow represents a specific electron-hole transition \(i\to a\) (e.g.\ A HOMO \(\to\) B LUMO), with its tail at the hole’s Wannier center and its head at the electron’s Wannier center, separated by $\Delta\mathbf{R}$. The number at each arrowhead denotes the percentage contribution of that transition to the overall exciton state, namely
\begin{equation}
    \mathcal{P}_{ai\Delta 
\mathbf{R}}^{S\mathbf{Q}}=\left|A_{a i \Delta \mathbf{R}}^{S \mathbf{Q}}\right|^2.
    \label{P_Rai}
\end{equation}

% \begin{figure}[!htb]
% \includegraphics[width=0.8\linewidth]{figures/arrow_naph_A.pdf}
% \caption{\label{fig:arrow_naph_A} Arrow plot visualization of exciton transitions in crystalline naphthalene (\(\mathbf{Q}=\mathbf{0}\)), showing only A site hole origins.  
% (a) Singlet, (b) triplet, and (c) top view of the \(ab\)‐plane projection.  Pink/blue dots mark A/B Wannier centers; arrow directions and numbers indicate \(\mathcal{P}_{ai\Delta\mathbf{R}}^{S\mathbf{Q}}\).}
% \end{figure}

Table~\ref{tab:acene_AB} summarizes the site‐resolved excitation weights and RMS distances, namely
\begin{subequations}
\begin{align}
   \mathcal{P}_{i \in \mathrm{X}}^{S \mathbf{Q}}&=\sum_{a, i \in \mathrm{X};\Delta \mathbf{R}}\left|A_{a i \Delta \mathbf{R}}^{S \mathbf{Q}}\right|^2 \label{P_AB}\\ \text{and }
   \sqrt{\braket{|\Delta\mathbf{R}|^2}}_{i \in \mathrm{X}}&=\sqrt{\frac{\sum_{a, i \in \mathrm{X}; \Delta\mathbf{R}} \left|A_{a i \Delta \mathbf{R}}^{S \mathbf{Q}}\right|^2\, |\Delta \mathbf{R}|^2}{\mathcal{P}_{i \in \mathrm{X}}^{S \mathbf{Q}}}},   
 \label{DR_AB}  
\end{align}
\end{subequations}
for $\mathrm{X}=\mathrm{A}$ or $\mathrm{B}$ at \(\mathbf{Q}=\mathbf{0}\), respectively. The symmetries of the four lattices are previously discussed in Sec. \ref{sec:com_detail}. Naphthalene and anthracene yield identical values on the A and B sublattices, reflecting their nonsymmorphic symmetry and consequent sublattice equivalence. However, tetracene and pentacene, which lack this symmetry, exhibit distinct A and B site values. The joint orbital-spatial resolution of our WFDX approach enables more accurate and detailed exciton post-processing and uncovers real-space symmetries concealed by other analyses.  As an illustration, we demonstrate how WFDX exposes the glide-mirror symmetry of the lattice, a property that remains indirect in conventional Bloch-function analyses.

\begin{table*}[!htb]
\caption{Site-resolved excitation weight and RMS electron–hole separation for the lowest \(\mathbf{Q}=\mathbf{0}\) singlet in acene crystals.}
\begin{ruledtabular}
\begin{tabular}{ccccc}
  & \multicolumn{2}{c}{Excitation weight (\%)} & \multicolumn{2}{c}{RMS distance $\sqrt{\langle|\Delta \mathbf{R}|^2\rangle}$ (\AA)} \\ \cmidrule(lr){2-3}\cmidrule(lr){4-5}
Number of rings & Hole on A & Hole on B  & Hole on A  & Hole on B \\ \midrule
2               & 50        & 50         & 2.66       & 2.66        \\ 
3               & 49        & 49         & 4.45       & 4.45        \\ 
4               & 59        & 40         & 5.34       & 5.52        \\ 
5               & 59        & 41         & 5.78       & 6.18        \\
\end{tabular}
\end{ruledtabular}
\label{tab:acene_AB}
\end{table*}

In Fig. \ref{fig:A_B_naph_arrow}, the yellow boxes highlight identical percentage contributions paired by glide-mirror symmetry in naphthalene, while the green boxes underscore their disparity in pentacene. This behavior follows directly from applying the nonsymmorphic glide-mirror operator \(P=\{\sigma|\tau\}\) to the exciton wavefunction in the Wannier basis (Eq.~\ref{eq:xct_wannier}), yielding:
\begin{equation}
\bigl\lvert A_{mn\Delta\mathbf{R}}^{S\mathbf{Q}}\bigr\rvert^2 = \bigl\lvert A_{\bar m\,\bar n,\,\sigma\Delta\mathbf{R}}^{S\mathbf{Q}}\bigr\rvert^2,  
\label{eq:eqaul_coeff}
\end{equation}
where $m,n \in \{ \mathrm{A}, \mathrm{B}\}$, $\bar m,\bar n$ are the complementary sublattice indices (e.g. if $m=\mathrm{A}$ then $\bar{m}=\mathrm{B}$). Thus each A\(\to\)B transition maps to its B\(\to\)A counterpart, with the separation vector reflected by the mirror operation \(\sigma\), which is exactly what we saw in Fig.~\ref{fig:A_B_naph_arrow}. This is equivalent to the more general statement that the magnitudes of the exciton amplitudes on the sublattice connected by nonsymmorphic symmetry are equal, and the electron-hole Wannier center difference is mapped from one sublattice to the other with the point group operation of the nonsymmorphic symmetry. A full derivation appears in Appendix~\ref{sec:nonsym}.

The nonsymmorphic symmetry may also be seen in reciprocal space, but is manisfested as a phase rather than a sublattice symmetry, as the single-particle Bloch function transforms as
\begin{equation}
P\,\psi_{n\mathbf{k}}(\mathbf{r}) = e^{i\mathbf{k}\cdot\boldsymbol{\tau}}\,\psi_{n,\,\sigma\mathbf{k}}(\mathbf{r}).  
\label{P_Bloch}
\end{equation}
Here $\boldsymbol{\tau}$ is the vector form of translation operation $\tau$, which enters only as a global phase, not as a coordinate shift.  

\section{\label{sec:dis}Discussion and Outlook}
The WFDX method introduced here adds both orbital and spatial resolution to the analysis of excitons in crystals, revealing sublattice symmetries obscured in the Bloch‐basis and in other existing methods.  Moreover, WFDX offers a direct route to study phase‐dependent excited‐state phenomena in real space.  For example, interaction‐driven topological effects in excitons are encoded in the phase of their envelope functions.  As noted in Ref.~\cite{davenport2024b}, prior work focusing solely on reciprocal‐space phases treats this information indirectly; more recently, there has been a push to understand such phase‐related properties, so‐called ``shift excitons", in a real‐space picture.  Drawing on the use of Wannier functions in single‐particle topology, such as the modern theory of polarization, where electronic Wannier centers shift under an applied field \cite{king-smith1993}, and the bulk photovoltaic shift current, in which light drives coherent displacements of charge centers \cite{ibanez-azpiroz2018}, WFDX promises to enable \textit{ab initio} studies of shift excitons in real materials.

We anticipate that WFDX arrow plots will serve as an efficient tool for real‐space exciton analysis in solids, complementing exciton correlation functions and NTOs much as band‐structure ``dot" plots for reciprocal‐space analysis.  At its core, WFDX rotates the exciton expansion from delocalized Bloch functions into a localized Wannier basis.  Similar Wannier‐based strategies have already accelerated excited‐state calculations \cite{marsili2017,marrazzo2024}, including implementations of the BSE in an electron-hole Wannier‐product basis \cite{merkel2023}, and we similarly expect WFDX to open new avenues for computing excitonic properties.

One immediate application is fine‐grid interpolation of exciton amplitudes.  By solving the BSE on a coarse mesh \(\mathbf{k}_\mathrm{co}\) to obtain \(A_{cv\mathbf{k}_\mathrm{co}}^{S\mathbf Q}\) and converting to real space via Eq.~\ref{eq:A_coeff_wannier} to extract $A_{ai\Delta\bar{\mathbf R}}^{S\mathbf Q}$, one can exploit the rapid decay of these real‐space coefficients to reconstruct amplitudes on any fine mesh \(\mathbf{k}_\mathrm{fi}\):
\begin{equation}
\begin{aligned}
&A_{cv\mathbf{k}_\mathrm{fi}}^{S\mathbf Q}
\\
&=\frac{1}{\sqrt{N_{\Delta \bar{\mathbf{R}}}}}
\sum_{ai \Delta\bar{\mathbf R}}^{N_{\mathcal{W}}}
e^{-i\,\mathbf{k}_{\mathrm{fi}}\cdot\Delta\bar{\mathbf R}}\,
U_{ca}(\mathbf{k}_{\mathrm{fi}}) \, A_{ai\Delta\bar{\mathbf R}}^{S\mathbf Q} \, U^{\dagger}_{iv}(\mathbf{k}_{\mathrm{fi}}-\mathbf{Q})\,,
\label{eq:iFFT}
\end{aligned}
\end{equation}
thereby greatly reducing computational cost. Here \(N_{\mathbf{\Delta \bar{\mathbf{R}}}}\) is the number of real-space points for the electron-hole Wannier center separation; in pracetice $N_{\mathbf{\Delta \bar{\mathbf{R}}}}=N_{\mathbf{k}}$ because the two grids are Fourier conjugates. Indeed, this procedure was employed to interpolate exciton coefficients in Fig.~\ref{fig:xct_k} across the bandstructure, demonstrating its practical effectiveness.

Beyond interpolation, WFDX also streamlines the evaluation of position‐operator–dependent properties. For example, it gives a direct, gauge‐invariant expression for the average electron–hole separation in Eq.~\ref{eq:eh-distance}. Another key observable is the ground‐to‐excited transition dipole, which in the conventional Bloch basis is written as
\begin{equation}
    \left\langle 0 \left| \hat{\mathbf{r}} \right|S\mathbf{Q} \right\rangle = \sum_{cv\mathbf{k}}A_{cv\mathbf{k}}^{S\mathbf{Q}} \, \left\langle v \, \mathbf{k}-\mathbf{Q}\left| \hat{\mathbf{r}} \right|c\mathbf{k} \right\rangle.
    \label{eq:dipole_k_0S}
\end{equation}
Because independent phase choices for valence and conduction Bloch states can arbitrarily shift this dipole’s phase, it is more convenient to express it in the Wannier basis, namely as
\begin{equation}
\begin{aligned}
    &\left\langle 0 \left| \hat{\mathbf{r}} \right|S\mathbf{Q} \right\rangle \\
    &= \frac{1}{\sqrt{N_{\bar{\mathbf{R}}}}} \sum_{\bar{\mathbf{R}}} e^{i \mathbf{Q} \cdot \bar{\mathbf{R}}} \left[ \sum_{a i \Delta \bar{\mathbf{R}}} A_{a i \Delta \bar{\mathbf{R}}}^{S \mathbf{Q}} \, \left \langle  i \bar{\mathbf{R}} \right | \hat{\mathbf{r}} \left | a \bar{\mathbf{R}}+\Delta \bar{\mathbf{R}} \right \rangle \right],
    \label{eq:dipole_R_0S}       
\end{aligned} 
\end{equation}
which is inherently gauge‐invariant, guaranteeing consistent and unambiguous dipole evaluations. Here $N_{\bar{\mathbf{R}}}$ is the number of real-space lattice vectors (Fourier-conjugate to the \textbf{k}-mesh), so in practice $N_{\bar{\mathbf{R}}}=N_{\mathbf{k}}$. In practice, the dipole approximation assumes long-wavelength light with negligible crystal momentum, it is standard to evaluate Eq.~\ref{eq:dipole_k_0S} and \ref{eq:dipole_R_0S} only in the  $\mathbf{Q} \rightarrow \mathbf{0}$ limit. Detailed derivations appear in Appendix ~\ref{sec:dipole_Wannier}.

\section{\label{sec:conclu}Conclusions}

In this work, we introduced the Wannier function decomposition of excitons (WFDX) method as a unified, real‐space framework for analysis of excitons in periodic solids. By mapping the $GW$–BSE exciton expansion in products of extended Bloch functions onto products of maximally localized Wannier functions, WFDX achieves high‐fidelity orbital and spatial decomposition at modest computational cost.  We demonstrated its effectiveness on herringbone-packed acene crystals, uncovering systematic trends in exciton delocalization with increasing ring number, differing spin multiplicities, and varying center‐of‐mass momentum.  Moreover, the combined orbital-spatial resolution afforded by WFDX reveals nonsymmorphic symmetry effects that remain obscured in reciprocal‐space treatments.  As discussed in Sec.~\ref{sec:dis}, WFDX promises to be a powerful tool for both analyzing and computing excited‐state properties in solids.  We envisage that it will become a standard complement to natural transition orbitals and exciton correlation functions, enabling more efficient post-processing and deeper insight into excited‐state phenomena across a broad range of semiconducting and insulating materials.  

\begin{acknowledgments}
This work was supported by the Center for Computational Study of Excited-State Phenomena in Energy Materials (C2SEPEM) at the Lawrence Berkeley National Laboratory, funded by the U.S. Department of Energy, Office of Science, Basic Energy Sciences, Materials Sciences and Engineering Division, under Contract No. DE-AC02-05CH11231. Computational resources were provided by the National Energy Research Scientific Computing Center (NERSC).
% J.B.H. and Z.T. contributed equally to this work.
\end{acknowledgments}

\appendix
\onecolumngrid

% \section{Appendix}
\section{Definitions and Conventions}
\subsection{\label{sec:coord}Center-of-Mass and Relative Coordinates for Exciton States}
For an exciton with COM momentum $\mathbf{Q}$, one enforces 
\(
\mathbf{k}_e - \mathbf{k}_h = \mathbf{Q},
\,
\mathbf{k}_e = \mathbf{k},
\,
\mathbf{k}_h = \mathbf{k} - \mathbf{Q}.
\)
The combined Bloch phases can then be recast as
\(
\mathbf{k}_e\!\cdot\!\mathbf{r}_e - \mathbf{k}_h\!\cdot\!\mathbf{r}_h
= \mathbf{Q}\!\cdot\!\mathbf{R}
+ \mathbf{k}_{\mathrm{rel}}\!\cdot\!\mathbf{r}_{\mathrm{rel}},
\)
where
\(
\mathbf{R} = \alpha\,\mathbf{r}_e + \beta\,\mathbf{r}_h,
\,
\mathbf{r}_{\mathrm{rel}} = \mathbf{r}_e - \mathbf{r}_h,
\,
\mathbf{k}_{\mathrm{rel}} = \beta\,\mathbf{k}_e + \alpha\,\mathbf{k}_h,
\)
with $\alpha+\beta=1$ determined by the electron and hole effective masses.  Here $\mathbf{R}$ is the exciton COM coordinate and $\mathbf{r}_{\mathrm{rel}}$ the relative coordinate \cite{haber2023}; correspondingly, $\mathbf{Q}$ is the COM momentum and $\mathbf{k}_{\text{rel}}$ is the relative momentum.

\subsection{\label{sec:coefficient_k2R}Transforming Exciton Coefficients from Reciprocal to Real Space}

We now convert the exciton expansion from the delocalized single-particle Bloch electron–hole product basis to the localized single-particle Wannier electron–hole product basis. Specifically,
\begin{equation}
\begin{aligned}
&\Psi_{S\mathbf{Q}}(\mathbf{r}_e,\mathbf{r}_h) \\
&= \sum_{cv\mathbf{k}}A_{cv\mathbf{k}}^{S\mathbf{Q}} \left[ \frac{1}{\sqrt{N_{\mathbf{k}}}} \, \sum_{a \bar{\mathbf{R}}+\Delta\bar{\mathbf{R}}}^{N_{\mathcal{W}}} e^{i\mathbf{k}\cdot (\bar{\mathbf{R}}+\Delta\bar{\mathbf{R}})}\, U^{\dagger}_{ac}(\mathbf{k})\, w_{a \bar{\mathbf{R}}+\Delta\bar{\mathbf{R}}}(\mathbf{r}_e)\right] \left[\frac{1}{\sqrt{N_{\mathbf{k}}}}\sum_{i \bar{\mathbf{R}}}^{N_{\mathcal{W}}} e^{-i(\mathbf{k}-\mathbf{Q})\cdot \bar{\mathbf{R}}}\, [U^{\dagger}_{iv}(\mathbf{k-Q})]^{\star}\, w_{i \bar{\mathbf{R}}}^{\star}(\mathbf{r}_h) \right]\\
&= \frac{1}{\sqrt{N_{\mathbf{k}}}} \sum_{\bar{\mathbf{R}}} e^{i \mathbf{Q}\cdot \bar{\mathbf{R}}} \sum_{ai\Delta \bar{\mathbf{R}}} \underbrace{\left[ \frac{1}{\sqrt{N_{\mathbf{k}}}} \sum_{cv\mathbf{k}}^{N_{\mathcal{W}}} e^{i\mathbf{k} \cdot \Delta \bar{\mathbf{R}}} \, U^{\dagger}_{ac}(\mathbf{k}) \, A_{cv\mathbf{k}}^{S\mathbf{Q}} \, U_{vi}(\mathbf{k-Q})\right]}_{A_{ai\Delta{\bar{\mathbf{R}}}}^{S\mathbf{Q}}} w_{a \bar{\mathbf{R}}+\Delta\bar{\mathbf{R}}}(\mathbf{r}_e) \, w_{i \bar{\mathbf{R}}}^{\star}(\mathbf{r}_h)\\
&= \frac{1}{\sqrt{N_{\mathbf{k}}}} \sum_{\bar{\mathbf{R}}} e^{i \mathbf{Q}\cdot \bar{\mathbf{R}}} \left[\sum_{ai\Delta \bar{\mathbf{R}}} A_{ai\Delta{\bar{\mathbf{R}}}}^{S\mathbf{Q}} \, w_{a \bar{\mathbf{R}}+\Delta\bar{\mathbf{R}}}(\mathbf{r}_e) \, w_{i \bar{\mathbf{R}}}^{\star}(\mathbf{r}_h)\right]. \label{eq:trans_Ak2Ar}
\end{aligned}
\end{equation}
The final line shows explicitly that the exciton is expanded in the localized electron–hole Wannier product basis (terms in brackets), while the outermost sum over $\bar{\mathbf{R}}$ with phase $e^{i \mathbf{Q}\cdot \bar{\mathbf{R}}}$ restores the Bloch character of the exciton at COM momentum $\mathbf{Q}$. The coefficients $A_{ai\Delta{\bar{\mathbf{R}}}}^{S\mathbf{Q}}$ are the real-space exciton amplitudes used throughout; their normalization is verified in Sec.~\ref{sec:norm_real_coeff}.

% In practice, we use a single unitary matrix $U(\mathbf{k})$ that rotates the entire composite manifold (valence + conduction) into the Wannier gauge; the electron and hole subspaces need not be treated separately. The real-space exciton coefficients can then be written in the compact form
% \begin{equation}
%     A_{mm' \Delta \bar{\mathbf{R}}}^{S \mathbf{Q}} = \frac{1}{\sqrt{N_{\mathbf{k}}}} \sum_{nn'\mathbf{k}}^{N_{\mathcal{W}}} e^{i\mathbf{k} \cdot \Delta \bar{\mathbf{R}}} \, U^{\dagger}_{mn}(\mathbf{k}) \, A_{nn'\mathbf{k}}^{S\mathbf{Q}} \, U_{n'm'}(\mathbf{k-Q}),\label{eq:A_r_nm}
% \end{equation}
% where $m, m'$ label Wannier orbitals, and $n, n'$ label Bloch bands in the chosen composite subspace. $A_{nn'\mathbf{k}}^{S\mathbf{Q}}$ is the zero-padded CV-block embedding of the usual TDA amplitudes: $A_{nn'\mathbf{k}}^{S\mathbf{Q}}=A_{cv\mathbf{k}}^{S\mathbf{Q}}$ for $n \in C \text{(conducation)}$, $n' \in V \text{(valence)}$, and zero otherwise.

\subsection{\label{sec:dipole_Wannier}Transition Dipole in Wannier Basis}

The ground-to-excited transition dipole can be recast in the Wannier basis by inserting the Bloch–Wannier transformations (Eq.~\ref{eq:Bloch-WF}) in Eq.~\ref{eq:dipole_k_0S},
\begin{equation}
    \begin{aligned}
        &\left\langle 0 \left| \hat{\mathbf{r}} \right|S\mathbf{Q} \right\rangle \\
        &= \sum_{cv\mathbf{k}}A_{cv\mathbf{k}}^{S\mathbf{Q}} \left( \frac{1}{\sqrt{N_{\mathbf{k}}}}\sum_{i\bar{\mathbf{R}}} e^{-i(\mathbf{k}-\mathbf{Q})\cdot \bar{\mathbf{R}}} U_{vi}(\mathbf{k}-\mathbf{Q}) \left \langle  i \bar{\mathbf{R}} \right |\right) \mathbf{r} 
        \left(  \frac{1}{\sqrt{N_{\mathbf{k}}}} \sum_{a\bar{\mathbf{R}}+\Delta \bar{\mathbf{R}}} e^{i\mathbf{k} \cdot (\bar{\mathbf{R}}+\Delta \bar{\mathbf{R}})} U_{ac}^{\dagger}(\mathbf{k}) \left |  a \bar{\mathbf{R}}+\Delta \bar{\mathbf{R}} \right \rangle \right)\\
        &= \frac{1}{\sqrt{N_{\mathbf{k}}}} \sum_{\bar{\mathbf{R}}} e^{i\mathbf{Q} \cdot \bar{\mathbf{R}}} \sum_{ai\Delta \bar{\mathbf{R}}}\underbrace{\left[ \frac{1}{\sqrt{N_{\mathbf{k}}}} \sum_{cv\mathbf{k}}^{N_{\mathcal{W}}} e^{i\mathbf{k} \cdot \Delta \bar{\mathbf{R}}} \, U^{\dagger}_{ac}(\mathbf{k}) \, A_{cv\mathbf{k}}^{S\mathbf{Q}} \, U_{vi}(\mathbf{k-Q})\right]}_{A_{ai\Delta{\bar{\mathbf{R}}}}^{S\mathbf{Q}}} \left \langle  i \bar{\mathbf{R}} \right | \hat{\mathbf{r}} \left | a \bar{\mathbf{R}}+\Delta \bar{\mathbf{R}} \right \rangle\\
        &=\frac{1}{\sqrt{N_{\bar{\mathbf{R}}}}} \sum_{\bar{\mathbf{R}}} e^{i \mathbf{Q} \cdot \bar{\mathbf{R}}} \left[ \sum_{a i \Delta \bar{\mathbf{R}}} A_{a i \Delta \bar{\mathbf{R}}}^{S \mathbf{Q}} \, \left \langle  i \bar{\mathbf{R}} \right | \hat{\mathbf{r}} \left | a \bar{\mathbf{R}}+\Delta \bar{\mathbf{R}} \right \rangle \right].
    \end{aligned}
\end{equation}
The algebra and final structure mirror Eq.~\ref{eq:trans_Ak2Ar}; the underbraced term on the second line is precisely the real-space (Wannier-basis) exciton amplitude $A_{a i \Delta \bar{\mathbf{R}}}^{S \mathbf{Q}}$.

\subsection{\label{sec:norm}Normalization}
\subsubsection{\label{sec:bloch-wannier}Single-particle Bloch and Wannier functions}

We adopt Born-von Karman boundary conditions with $N_{\mathbf{k}}$ $\mathbf{k}$-points (equivalently, $N_{\mathbf{k}}$ unit cells). The single-particle Bloch states are written as
\begin{equation}
    \psi_{n\mathbf{k}}=\frac{1}{\sqrt{N_{\mathbf{k}}}}e^{i\mathbf{k}\cdot \mathbf{r}} u_{n\mathbf{k}}(\mathbf{r}),\label{eq:conv_Bloch_single}
\end{equation}
where the cell-periodic part $u_{n\mathbf{k}}(\mathbf{r})$ are orthonormal in the unit cell of volume $V_{\mathrm{uc}}$, namely
\begin{equation}
    \int_{V_{\mathrm{uc}}} d \mathbf{r} \, u_{n \mathbf{k}}^{\star}(\mathbf{r}) u_{n^{\prime} \mathbf{k}}(\mathbf{r})=\delta_{n n^{\prime}}.
\end{equation}
With this convention the Bloch functions $\psi_{n\mathbf{k}}(\mathbf{r})$ are orthonormal over the crystal volume $V_{\mathbf{k}}=V_{\mathrm{uc}} N_{\mathbf{k}}$:
\begin{equation}
    \int_{V_{\mathbf{k}}} d \mathbf{r} \, \psi_{n \mathbf{k}}^{\star}(\mathbf{r}) \psi_{n^{\prime} \mathbf{k}^{\prime}}(\mathbf{r})=\delta_{\mathbf{k k}^{\prime}} \delta_{n n^{\prime}}.
\end{equation}

Using Eq.~\ref{eq:WF-Bloch} and the Bloch orthonormality, the Wannier functions are also orthomormal over the crystal volume $V_{\bar{\mathbf{R}}}=V_{\mathbf{k}}$:
\begin{equation}
    \begin{aligned}
       & \int_{V_{\bar{\mathbf{R}}}} d\mathbf{r} \, w_{m \bar{\mathbf{R}}}^{\star}(\mathbf{r}) \, w_{m' \bar{\mathbf{R}}}(\mathbf{r})\\
       &=\int_{V_{\bar{\mathbf{R}}}} d\mathbf{r} \, \frac{1}{N_{\mathbf{k}}} \left(\sum_{n\mathbf{k}}^{N_{\mathcal{W}}} e^{-i\mathbf{k}\cdot\bar{\mathbf{R}}} U_{nm}(\mathbf{k}) \psi_{n\mathbf{k}}(\mathbf{r})\right) \left(\sum_{n'\mathbf{k}'}^{N_{\mathcal{W}}} e^{i\mathbf{k}'\cdot\bar{\mathbf{R}}'} U^{\dagger}_{m'n'}(\mathbf{k}') \psi_{n'\mathbf{k}'}^{\star}(\mathbf{r})\right)\\
       &=\frac{1}{N_{\mathbf{k}}} \sum_{nn'\mathbf{k}\mathbf{k}'}^{N_{\mathcal{W}}}e^{i(\mathbf{k}'\cdot \bar{\mathbf{R}}'-\mathbf{k} \cdot \bar{\mathbf{R}})}U^{\dagger}_{m'n'}(\mathbf{k}')U_{nm}(\mathbf{k})\underbrace{\left[ \int_{V_{\bar{\mathbf{R}}}} d\mathbf{r} \, \psi_{n \mathbf{k}}^{\star}(\mathbf{r}) \psi_{n^{\prime} \mathbf{k}^{\prime}}(\mathbf{r}) \right]}_{\delta_{\mathbf{k k}^{\prime}} \delta_{n n^{\prime}}}\\
       &= \frac{1}{N_{\mathbf{k}}} \sum_{n\mathbf{k}}^{N_{\mathcal{W}}} \, e^{i\mathbf{k}\cdot (\bar{\mathbf{R}}'- \bar{\mathbf{R}})} \, U^{\dagger}_{m'n}(\mathbf{k})\,U_{nm}(\mathbf{k})=\frac{1}{\cancel{N_{\mathbf{k}}}} \cancel{N_{\mathbf{k}}}\, \delta_{\bar{\mathbf{R}}\bar{\mathbf{R}}'} \left[ U^{\dagger}(\mathbf{k}) \, U(\mathbf{k})\right]_{mm'}\\
       &=\delta_{\bar{\mathbf{R}}\bar{\mathbf{R}}'}\delta_{m m'}.
    \end{aligned}
\end{equation}

% \begin{equation}
% \begin{aligned}
% \psi_{n\mathbf{k}}(\mathbf{r})
% &=\frac{1}{\sqrt{N_{\mathbf{k}}}}
% \sum_{m \bar{\mathbf{R}}}^{N_{\mathcal{W}}}
% e^{\,i\mathbf{k}\cdot\bar{\mathbf{R}}}\,
% U^{\dagger}_{mn}(\mathbf{k})
% \Biggl[
%   \frac{1}{\sqrt{N_{\mathbf{k}}}}
%   \sum_{p \mathbf{k}'}^{N_{\mathcal{W}}}
%     e^{-\,i\mathbf{k}'\cdot\bar{\mathbf{R}}}\,
%     U_{p m}(\mathbf{k}')\,
%     \psi_{p\mathbf{k}'}(\mathbf{r})
% \Biggr] \\[8pt]
% %
% &=\frac{1}{N_{\mathbf{k}}}
% \sum_{p \mathbf{k}' m \bar{\mathbf{R}}}^{N_{\mathcal{W}}}
%   e^{\,i(\mathbf{k}-\mathbf{k}')\cdot\bar{\mathbf{R}}}\,
%   U^{\dagger}_{mn}(\mathbf{k})
%   \,U_{p m}(\mathbf{k}')\,
%   \psi_{p\mathbf{k}'}(\mathbf{r}) \\[8pt]
% %
% &=\frac{1}{N_{\mathbf{k}}}
% \sum_{p \mathbf{k}' m}^{N_{\mathcal{W}}}
%   \Bigl[\sum_{\bar{\mathbf{R}}}
%     e^{\,i(\mathbf{k}-\mathbf{k}')\cdot\bar{\mathbf{R}}}\Bigr]
%   \;U^{\dagger}_{mn}(\mathbf{k})
%   \,U_{p m}(\mathbf{k}')\,
%   \psi_{p\mathbf{k}'}(\mathbf{r}) \\[8pt]
% %
% &=\frac{1}{\cancel{N_{\mathbf{k}}}}
% \sum_{p \mathbf{k}' m}^{N_{\mathcal{W}}}
%   \underbrace{\cancel{N_{\bar{\mathbf{R}}}}}_{N_{\mathbf{k}}}\,\delta_{\mathbf{k} \mathbf{k}'}
%   \;U^{\dagger}_{mn}(\mathbf{k})
%   \,U_{p m}(\mathbf{k}')
%   \,\psi_{p\mathbf{k}'}(\mathbf{r}) \\[8pt]
% %
% &=\sum_p \left[ \sum_{m}^{N_{\mathcal{W}}}
%   U_{p m}(\mathbf{k}) \, U^{\dagger}_{mn}(\mathbf{k})\right]
%   \,\psi_{p\mathbf{k}}(\mathbf{r}) \\[6pt]
% %
% &=\sum_{p}
%   \delta_{np}\,
%   \psi_{p\mathbf{k}}(\mathbf{r})
%   = \psi_{n\mathbf{k}}(\mathbf{r}) \,.
% \end{aligned}
% \end{equation}

\subsubsection{\label{sec:norm_real_coeff}Exciton wave function in Wannier basis}

With the Bloch state normalization of Eq.~\ref{eq:conv_Bloch_single}, the exciton eigenvectors are orthonormal in the Bloch product basis. We now show that the real-space (Wannier-basis) amplitudes are likewise orthonormal, namely
\begin{equation}
    \begin{aligned}
        &\sum_{ai\Delta \bar{\mathbf{R}}} A_{ai\Delta \bar{\mathbf{R}}}^{S\mathbf{Q}\, \star} A_{ai\Delta \bar{\mathbf{R}}}^{S'\mathbf{Q}'} \\
        &= \sum_{ai\Delta \bar{\mathbf{R}}} \left(\frac{1}{\sqrt{N_{\mathbf{k}}}} \sum_{cv\mathbf{k}}^{N_{\mathcal{W}}} e^{-i\mathbf{k} \cdot \Delta \bar{\mathbf{R}}} \left[ U^{\dagger}_{ac}(\mathbf{k}) \right]^{\star} \, A_{cv\mathbf{k}}^{S\mathbf{Q} \star} \, U_{vi}^{\star}(\mathbf{k-Q})  \right)  \left(\frac{1}{\sqrt{N_{\mathbf{k}}}} \sum_{c'v'\mathbf{k}'}^{N_{\mathcal{W}}} e^{i\mathbf{k}' \cdot \Delta \bar{\mathbf{R}}} U^{\dagger}_{ac'}(\mathbf{k}') \, A_{c'v'\mathbf{k}'}^{S'\mathbf{Q}'}U_{v'i}(\mathbf{k'-Q'}) \right)\\
        &= \frac{1}{N_{\mathbf{k}}} \sum_{\Delta \bar{\mathbf{R}}\, \mathbf{k} \, \mathbf{k}'} e^{i \Delta \bar{\mathbf{R}} \cdot (\mathbf{k}' - \mathbf{k})} \sum_{cvc'v'} \left[ \sum_{a}^{N_{\mathcal{W}}} U_{ca}(\mathbf{k}) \, U^{\dagger}_{ac'}(\mathbf{k'}) \right] A_{cv\mathbf{k}}^{S\mathbf{Q}\star} A_{c'v'\mathbf{k'}}^{S'\mathbf{Q}'} \,  \left[ \sum_{i}^{N_{\mathcal{W}}} U_{v'i}(\mathbf{k}'-\mathbf{Q}') \, U^{\dagger}_{iv}(\mathbf{k}-\mathbf{Q}) \right] \\
        &= \frac{1}{\cancel{N_{\mathbf{k}}}} \sum_{\mathbf{k} \mathbf{k}'}  \underbrace{\cancel{N_{\Delta \bar{\mathbf{R}}}}}_{N_{\mathbf{k}}} \delta_{\mathbf{k} \mathbf{k}'} \sum_{cvc'v'} \left[U(\mathbf{k}) U^{\dagger}(\mathbf{k'}) \right]_{cc'} A_{cv\mathbf{k}}^{S\mathbf{Q} \star} A_{c'v'\mathbf{k}'}^{S'\mathbf{Q}'} \left[ U(\mathbf{k}'-\mathbf{Q}') \, U^{\dagger}(\mathbf{k}-\mathbf{Q}) \right]_{v'v}\\
        &= \sum_{cc'} \delta_{c c'} \sum_{vv'} \delta_{v v'} \, \delta_{\mathbf{Q} \mathbf{Q}'} \sum_{\mathbf{k}} A_{cv\mathbf{k}}^{S\mathbf{Q} \star} A_{c'v'\mathbf{k}}^{S'\mathbf{Q}'} \\
        &=\delta_{\mathbf{Q} \mathbf{Q}'} \left(\sum_{cv\mathbf{k}} A_{cv\mathbf{k}}^{S\mathbf{Q} \star} A_{cv\mathbf{k}}^{S'\mathbf{Q}'}\right) = \delta_{S S'} \delta_{\mathbf{Q} \mathbf{Q}'}.
    \end{aligned}
\end{equation}

It immediately follows that exciton state expanded in the Wannier basis are also orthonormal:
\begin{equation}
    \begin{aligned}
        \big\langle S \mathbf{Q} | S'\mathbf{Q}' \big\rangle &= \frac{1}{N_{\mathbf{k}}} \int_{V_{\mathbf{k}}} d \mathbf{r}_e \int_{V_{\mathbf{k}}} d \mathbf{r}_h \left[ \sum_{\bar{\mathbf{R}}} e^{-i \mathbf{Q} \cdot \bar{\mathbf{R}}} \left( \sum_{ai\Delta \bar{\mathbf{R}}} A_{ai\Delta \bar{\mathbf{R}}}^{S\mathbf{Q}\star} w_{a\bar{\mathbf{R}}+\Delta \bar{\mathbf{R}}}^{\star}(\mathbf{r}_e)w_{i\bar{\mathbf{R}}}(\mathbf{r}_h)\right)\right]\\
        &\left[ \sum_{\bar{\mathbf{R}}'} e^{i \mathbf{Q}' \cdot \bar{\mathbf{R}}'} \left( \sum_{a'i'\Delta \bar{\mathbf{R}}'} A_{a'i'\Delta \bar{\mathbf{R}}'}^{S'\mathbf{Q}'\star} w_{a'\bar{\mathbf{R}}'+\Delta \bar{\mathbf{R}}'}(\mathbf{r}_e)w_{i'\bar{\mathbf{R}}'}^{\star}(\mathbf{r}_h)\right)\right]\\
        &= \frac{1}{N_{\mathbf{k}}} \sum_{\bar{\mathbf{R}}\bar{\mathbf{R}}'} e^{i (\mathbf{Q}' \cdot \bar{\mathbf{R}}'-\mathbf{Q} \cdot \bar{\mathbf{R}})} \sum_{aia'i'\Delta \bar{\mathbf{R}} \Delta \bar{\mathbf{R}}'} A_{ai\Delta \bar{\mathbf{R}}}^{S\mathbf{Q}\star} A_{a'i'\Delta \bar{\mathbf{R}}'}^{S'\mathbf{Q}'} \left[\int_{V_{\mathbf{k}}} d \mathbf{r}_e w_{a\bar{\mathbf{R}}+\Delta \bar{\mathbf{R}}}^{\star}(\mathbf{r}_e) \, w_{a'\bar{\mathbf{R}}'+\Delta \bar{\mathbf{R}}'}(\mathbf{r}_e) \right] \\
        &\left[\int_{V_{\mathbf{k}}} d \mathbf{r}_h w_{i\bar{\mathbf{R}}}(\mathbf{r}_h) \, w_{i'\bar{\mathbf{R}}'}^{\star}(\mathbf{r}_h) \right]\\
        &= \frac{1}{N_{\mathbf{k}}} \sum_{\bar{\mathbf{R}}\bar{\mathbf{R}}'} e^{i (\mathbf{Q}' \cdot \bar{\mathbf{R}}'-\mathbf{Q} \cdot \bar{\mathbf{R}})} \sum_{aia'i'\Delta \bar{\mathbf{R}} \Delta \bar{\mathbf{R}}'} A_{ai\Delta \bar{\mathbf{R}}}^{S\mathbf{Q}\star} A_{a'i'\Delta \bar{\mathbf{R}}'}^{S'\mathbf{Q}'} \, \delta_{aa'} \delta_{\bar{\mathbf{R}}\bar{\mathbf{R}}'} \, \delta_{\Delta \bar{\mathbf{R}} \Delta \bar{\mathbf{R}}'} \delta_{ii'}\\
        &= \frac{1}{N_{\mathbf{k}}} \sum_{\bar{\mathbf{R}}} e^{i (\mathbf{Q}' -\mathbf{Q}) \cdot \bar{\mathbf{R}}} \sum_{ai\Delta \bar{\mathbf{R}}} A_{ai\Delta \bar{\mathbf{R}}}^{S\mathbf{Q}\star} A_{ai\Delta \bar{\mathbf{R}}}^{S'\mathbf{Q}'}\\
        &= \frac{1}{\cancel{N_{\mathbf{k}}}} \underbrace{\cancel{N_{\bar{\mathbf{R}}}}}_{N_{\mathbf{k}}} \delta_{S S'} \delta_{\mathbf{Q}\mathbf{Q}'}= \delta_{S S'} \delta_{\mathbf{Q} \mathbf{Q}'}.
    \end{aligned}
\end{equation}
Consequently, expressing the exciton in the Wannier basis preserves orthonormality: the Wannier-basis exciton eigenvectors remain orthonormal exactly as in the Bloch basis.

\subsubsection{\label{sec:norm_k_fi_coeff}Fine grid reciprocal space coefficients}
We verify that the interpolation formula Eq.~\ref{eq:iFFT} produces properly normalized exciton coefficients at any $\mathbf{k}$-point and introduces no spurious phase. Inserting Eq.~\ref{eq:A_coeff_wannier} into Eq.~\ref{eq:iFFT} gives
\begin{equation}
\begin{aligned}
    A_{cv\mathbf{k}_{\text{fi}}}^{S\mathbf{Q}} &= \frac{1}{\sqrt{N_{\Delta \bar{\mathbf{R}}}}} \sum_{ai\Delta \bar{\mathbf{R}}}^{N_{\mathcal{W}}} e^{-i\mathbf{k}_{\text{fi}} \cdot \Delta \bar{\mathbf{R}}} \, U_{ca}(\mathbf{k}_{\text{fi}}) A_{ai\Delta \bar{\mathbf{R}}}^{S\mathbf{Q}}U^{\dagger}_{iv}(\mathbf{k}_{\text{fi}}-\mathbf{Q})\\
    &= \frac{1}{\sqrt{N_{\Delta \bar{\mathbf{R}}}}} \sum_{ai\Delta \bar{\mathbf{R}}}^{N_{\mathcal{W}}} e^{-i\mathbf{k}_{\text{fi}} \cdot \Delta \bar{\mathbf{R}}} \, U_{ca}(\mathbf{k}_{\text{fi}}) \left[ \frac{1}{\sqrt{N_{\mathbf{k}}}} \sum_{c'v'\mathbf{k}}^{N_{\mathcal{W}}} e^{i \mathbf{k} \cdot \Delta \bar{\mathbf{R}}}\, U^{\dagger}_{ac'}(\mathbf{k})\, A_{c'v'\mathbf{k}}^{S\mathbf{Q}} \, U_{v'i}(\mathbf{k}-\mathbf{Q})  \right]  U^{\dagger}_{iv}(\mathbf{k}_{\text{fi}}-\mathbf{Q})\\
    &= \frac{1}{\sqrt{N_{\Delta \bar{\mathbf{R}}}}} \frac{1}{\sqrt{N_{\mathbf{k}}}} \sum_{\mathbf{k}} \left[ \sum_{\Delta \bar{\mathbf{R}}} e^{i(\mathbf{k}-\mathbf{k}_{\text{fi}}) \cdot \Delta \bar{\mathbf{R}}}\right] \sum_{c'v'} \left[\sum_{a}^{N_{\mathcal{W}}} U_{ca}(\mathbf{k}_{\text{fi}})\,U^{\dagger}_{ac'}(\mathbf{k})\right] A_{c'v'\mathbf{k}}^{S\mathbf{Q}} \left[\sum_{i}^{N_{\mathcal{W}}} U_{v'i}(\mathbf{k}-\mathbf{Q}) U^{\dagger}_{iv}(\mathbf{k}_{\text{fi}}-\mathbf{Q}) \right] \\
    &= \frac{1}{\cancel{N_{\Delta \bar{\mathbf{R}}}}} \sum_{\mathbf{k}} \cancel{N_{\Delta \bar{\mathbf{R}}}} \delta_{\mathbf{k}\mathbf{k}_{\text{fi}}} \, \sum_{c'v'}  \left[ U(\mathbf{k}_{\text{fi}}) U^{\dagger}(\mathbf{k})\right]_{cc'} A_{c'v'\mathbf{k}}^{S\mathbf{Q}} \left[U(\mathbf{k}-\mathbf{Q})U^{\dagger}(\mathbf{k}_{\text{fi}}-\mathbf{Q}) \right]_{v'v}\\
    &= \sum_{c'v'} \delta_{cc'} A_{c'v'\mathbf{k}_{\text{fi}}}^{S\mathbf{Q}} \delta_{v'v} = A_{cv\mathbf{k}_{\text{fi}}}^{S\mathbf{Q}}.
\end{aligned}
\end{equation}
Hence no additional normalization factor or phase is introduced by the
interpolation.

% \section{\label{sec:MLWF}MLWF Construction and Band-Structure Characterization}
% To obtain MLWFs that faithfully reproduce isolated molecular orbitals, we define separate energy windows for each subspace of crossing bands, and perform wannierization in each using symmetry-adapted technique \cite{sakuma2013}; we then assemble the subspaces into the full manifold to construct $U$. 

\section{\label{sec:acene_band_MLWF}MLWF Construction and Band Structure Characterization}

To obtain MLWFs that faithfully reproduce isolated molecular orbitals, we define separate energy windows for each subspace of crossing bands, and perform wannierization in each using symmetry-adapted technique \cite{sakuma2013}; we then assemble the subspaces into the full manifold to construct $U$. 

Given the single-particle energy, $\varepsilon_{n\mathbf{k}}$, and the corresponding Wannier rotation matrices, $U_{nm}(\mathbf{k})$, on a uniform grid of $\mathbf{k}$-points, we first Wannier-Fourier transform to obtain hopping matrix elements
\begin{equation}
    \begin{aligned}
        H_{mm'}(\bar{\mathbf{R}})=\sum_{n\mathbf{k}} e^{-i\mathbf{k} \cdot \bar{\mathbf{R}}} U^{\dagger}_{mn}(\mathbf{k}) \, \varepsilon_{n\mathbf{k}} \, U_{nm'}(\mathbf{k}),     
    \end{aligned}
\end{equation}
and subsequently Fourier transform to an arbitrary $\mathbf{k}_{\text{fi}}$-point to find
\begin{equation}
    \begin{aligned}
        H_{mm'}(\mathbf{k}_{\text{fi}})=\sum_{\bar{\mathbf{R}}} e^{i\mathbf{k}_{\text{fi}} \cdot \bar{\mathbf{R}}} H_{mm'}(\bar{\mathbf{R}}).
    \end{aligned}
\end{equation}
Diagonalizing $H(\mathbf{k}_{\text{fi}})$ yields eigenvalues and eigenstates $\left | n \mathbf{k_\text{fi}} \right \rangle$ expanded in the Wannier basis $\left\{\left | m \bar{\mathbf{R}}\right \rangle \right \}$. Assigning a distinct color to each Wannier function, we construct a colormap from the weights $\left | \langle m\bar{\mathbf{R}}| n \mathbf{k_\text{fi}} \rangle \right |^2$, and plot the Wannier contributions to every Bloch state on the selected $\mathbf{k}$-path. In this work, we supply $GW$ quasiparticle energies as inputs to the interpolation. The $\mathbf{k}$-path is $L (-0.5, 0.5, 0) \rightarrow X(-0.5,0,0) \rightarrow \Gamma(0,0,0) \rightarrow Y(0,0.5,0) \rightarrow L(0.5,0.5,0)$, with coordinates given in fractional reciprocal-lattice units. 

\begin{figure}[H]
\centering
\includegraphics[width=0.85\linewidth]{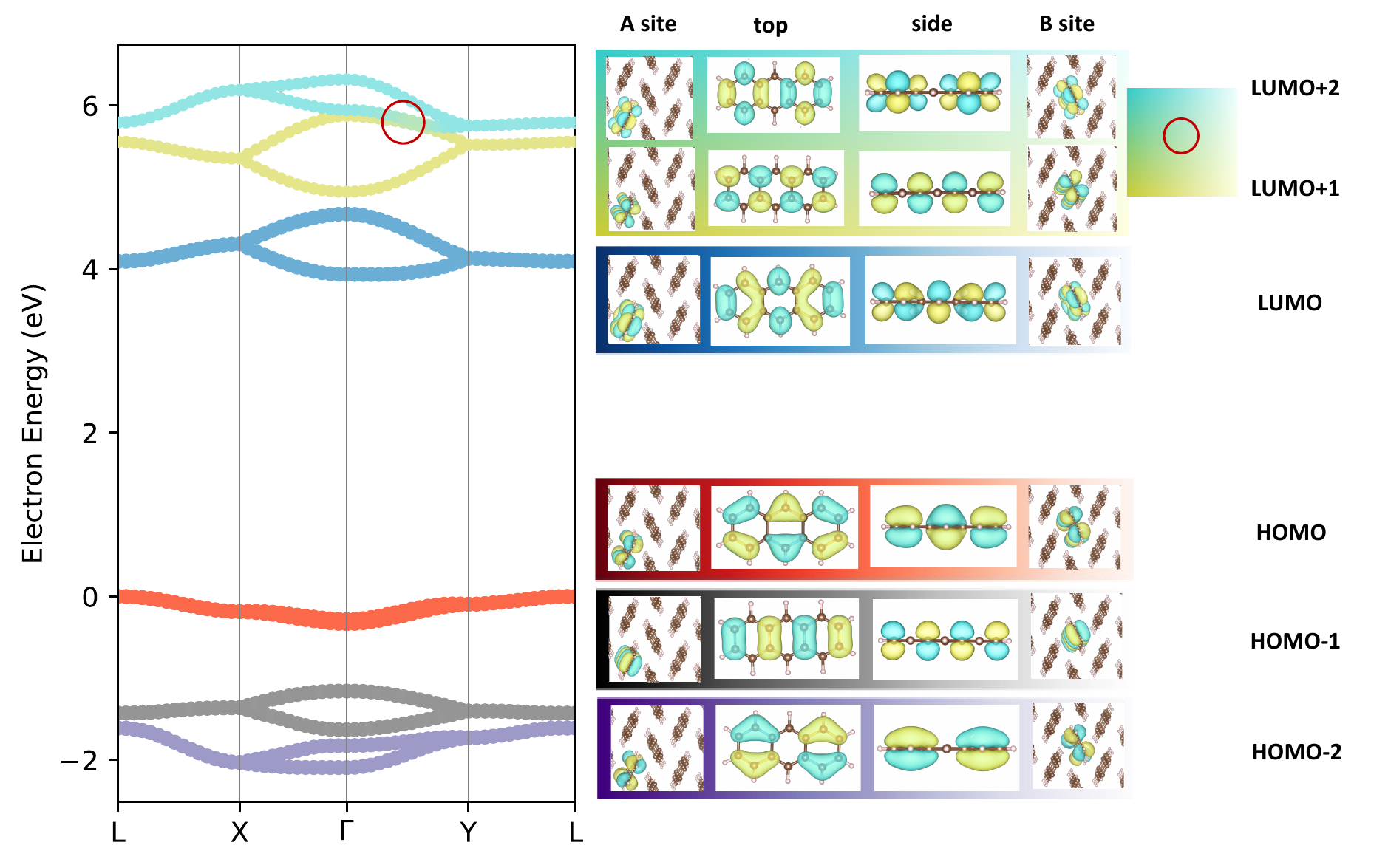}
\caption{\label{fig:anth_band_orbital} 
Band structure character of MLWFs in crystal anthracene. The near-midpoint colors across the bands indicate that each Bloch state is approximately an equal-weight superposition of MLWFs on the A and B sublattices. The red-circled region highlights mixing between LUMO+1 and LUMO+2; the rightside color bar quantifies their contributions.}
\end{figure}

\begin{figure}[H]
\centering
\includegraphics[width=0.9\linewidth]{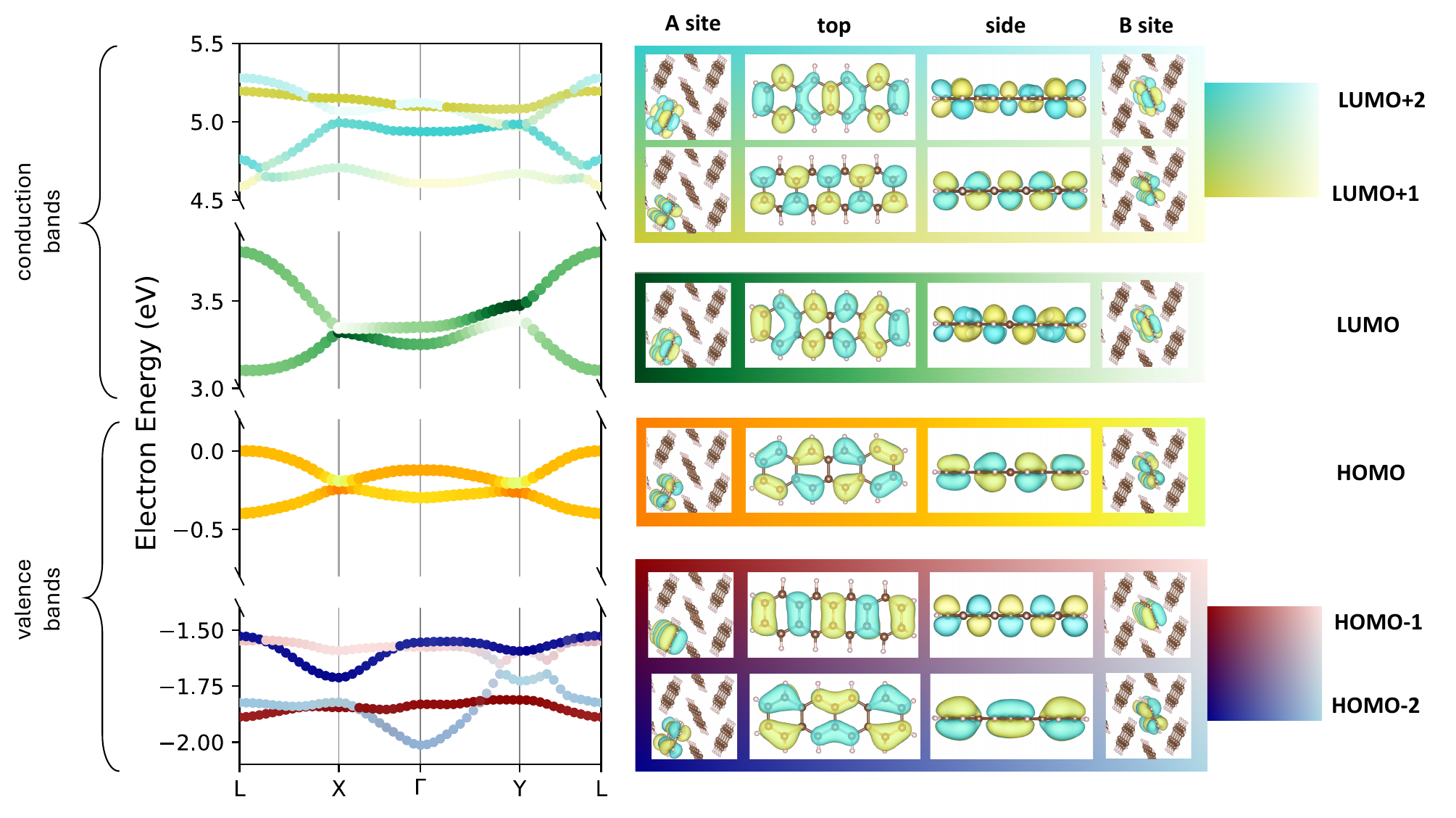}
\caption{\label{fig:tetra_band_orbital}Band structure character of MLWFs in crystal tetracene. The pronounced color variation indicates that each Bloch state comprises an inequivalent superposition of A- and B-sublattice Wannier functions, with clear mixing between HOMO-2 and HOMO-1 as well as between LUMO+1 and LUMO+2. For clarity, four vertically stacked, zoomed panels are shown for the bands forming HOMO-2/HOMO-1, HOMO, LUMO, and LUMO+1/LUMO+2, each with its own energy scale; the y-axis is broken between panels.}
\end{figure}

\begin{figure}[H]
\centering
\includegraphics[width=0.9\linewidth]{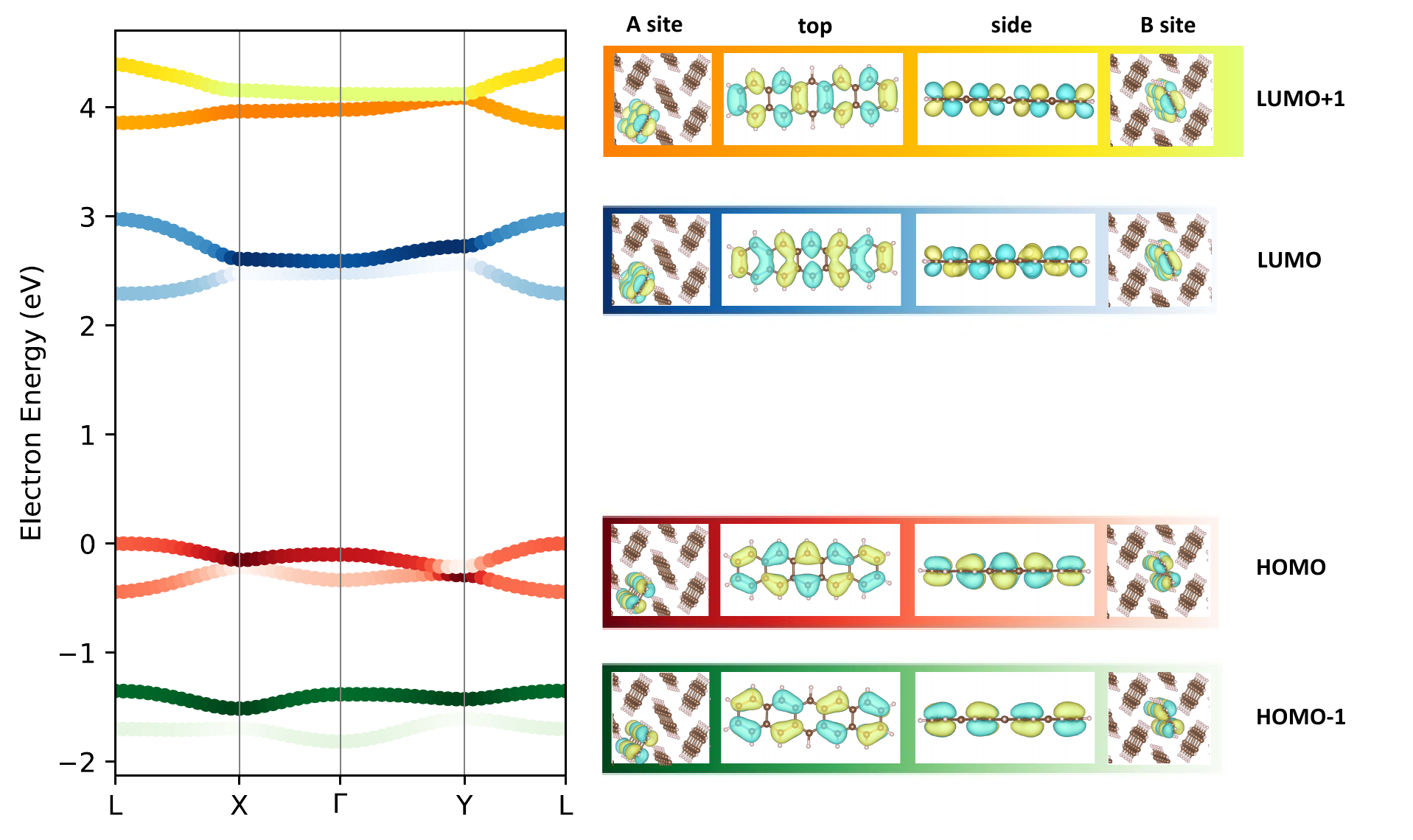}
\caption{\label{fig:penta_band_orbital}Band structure character of MLWFs in crystal pentacene.  The pronounced color variation indicates that each Bloch state comprises an inequivalent superposition of A- and B-sublattice Wannier functions.}
\end{figure}

Fig.~\ref{fig:anth_band_orbital} - \ref{fig:penta_band_orbital}, together with Fig.~\ref{fig:naph_band_orbital} in the main text, present the interpolated $GW$ band structures and their character of MLWFs. Left: $GW$ quasiparticle band structure of crystalline acenes, referenced to the valence band maximum (VBM set to 0 eV). Curves are colored by MLWF sublattice character (dark: site A; light: site B). Right: Isosurfaces of the MLWFs for (HOMO–2,) HOMO–1, HOMO, LUMO, LUMO+1 (and LUMO+2), columns from left to right show A site, top, side and B site views.

% \begin{figure}[!htb]

\section{\label{sec:acene_coupl}Intermolecular Coupling in Acene Crystals}
As shown in Table~\ref{tab:coupling}, the bandwidths of the two highest valence bands (forming HOMO-like Wannier functions) increase with the number of rings in the acenes, indicating stronger intermolecular coupling.

\begin{table}[!htb]
\caption{Bandwidth of the two highest valence bands in acene crystals.}
\begin{ruledtabular}
\begin{tabular}{cc}
Number of rings              & Bandwidth (meV)  \\ \midrule
2              & 28     \\ 
3              & 54     \\ 
4              & 505    \\ 
5              & 507    \\            
\end{tabular}
\end{ruledtabular}
\label{tab:coupling}
\end{table}

\section{\label{sec:Q_xct_dispersion} COM Dependence of the Exciton Spatial Character}
 In practice, we identified the smoothest connection of exciton bands by maximizing the overlap between exciton wave functions at adjacent \textbf{Q}-points (which sometimes required denser sampling along the \textbf{Q}-path) \cite{haber2023}. At each step, the state $S$ at $\mathbf{Q}$ is connected with the state $S'$ at the next $\mathbf{Q}'$ having the largest overlap, yielding a continuous tracking along the path. The resulting continuous, often monotonic, evolution of the exciton spatial character provides an internal consistency check on the assignment. Small deviations from monotonic behavior observed for the third and fourth singlet branches in tetracene and pentacene may be physical or may stem from incomplete convergence of the exciton dispersion. Achieving fully converged dispersions in acene crystals is nontrivial; a systematic convergence study is left to future work. 
 % \ZT{Jonah, could you also comment on this, it is another scientific related thing I am not sure. I think since the deviation happens at the Q point close to $\Gamma$, it might be related to LT splitting. The nearest neighbour overlap is more than 60\%, but when I include more bands for BSE absorp solvation, the overlap changes a lot. It can assign totally different connection way, even if both of them looks dominant.}

\begin{figure}[H]
\includegraphics[width=1\linewidth]{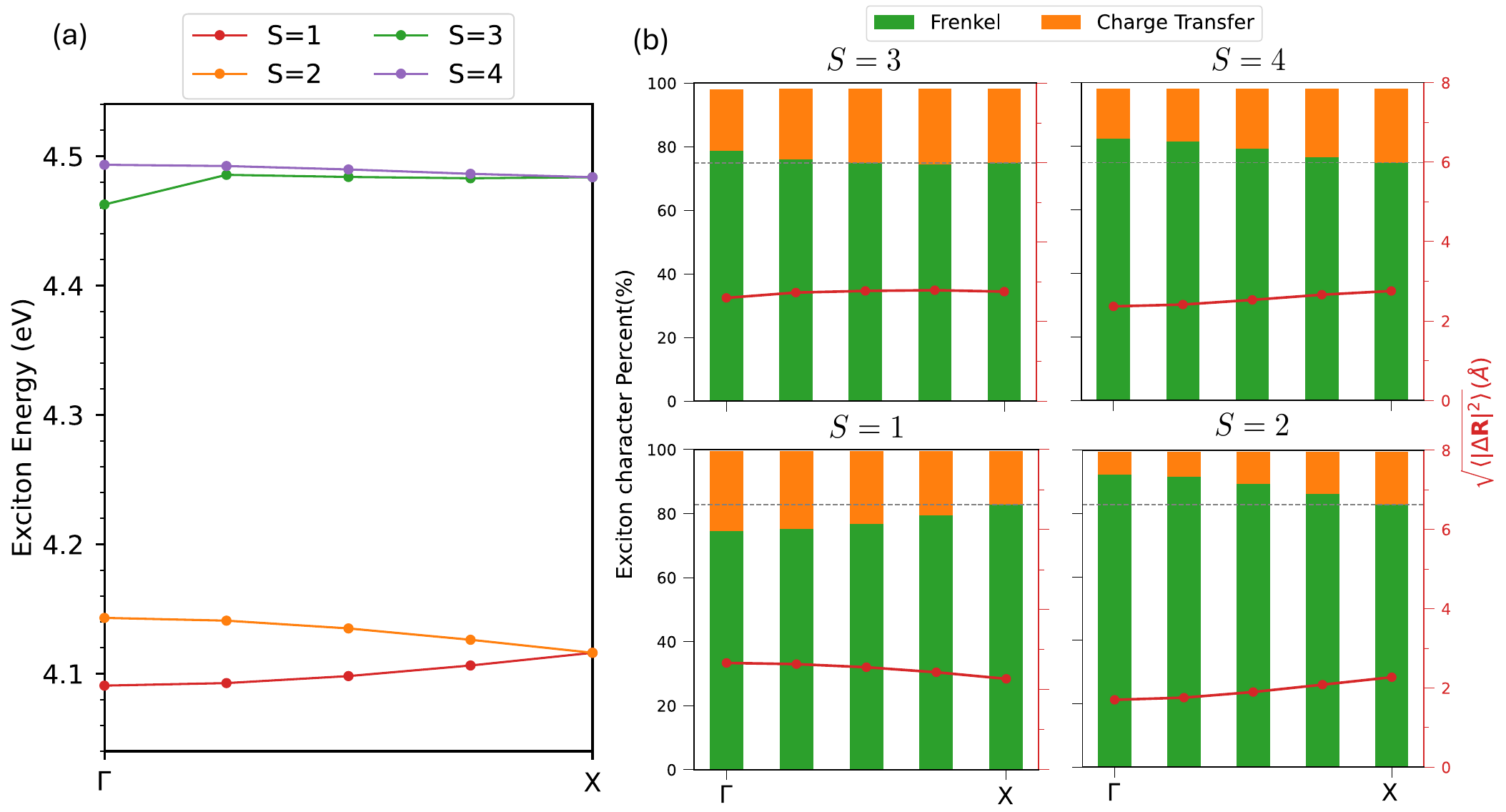}
\caption{\label{fig:naph_Q_S}$\mathbf{Q}$-dependence of singlet exciton spatial character in crystal naphthalene.}
\end{figure}

\begin{figure}[H]
\includegraphics[width=1\linewidth]{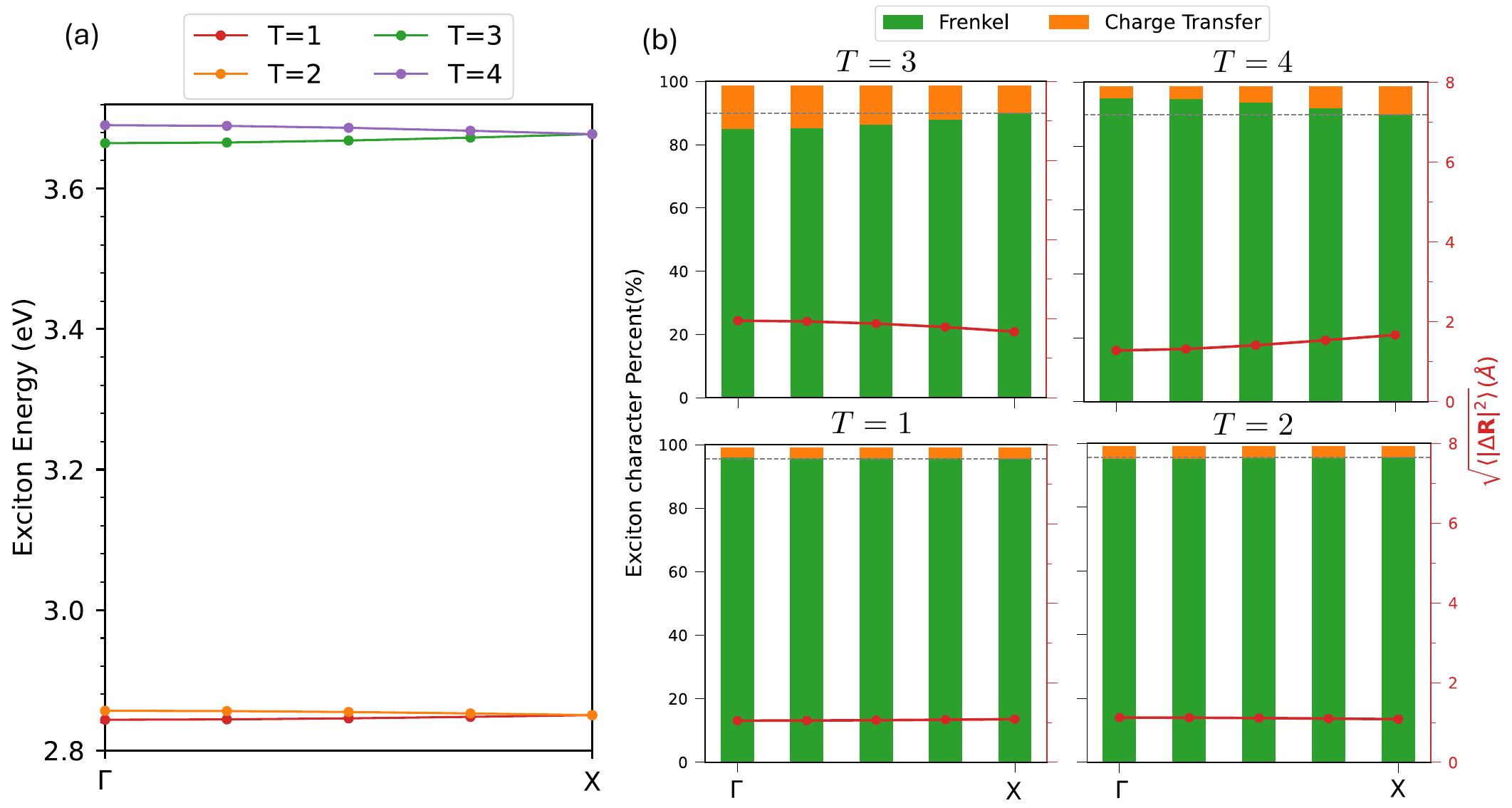}
\caption{\label{fig:naph_Q_T}$\mathbf{Q}$-dependence of triplet exciton spatial character in crystal naphthalene.}
\end{figure}

\begin{figure}[H]
\includegraphics[width=1\linewidth]{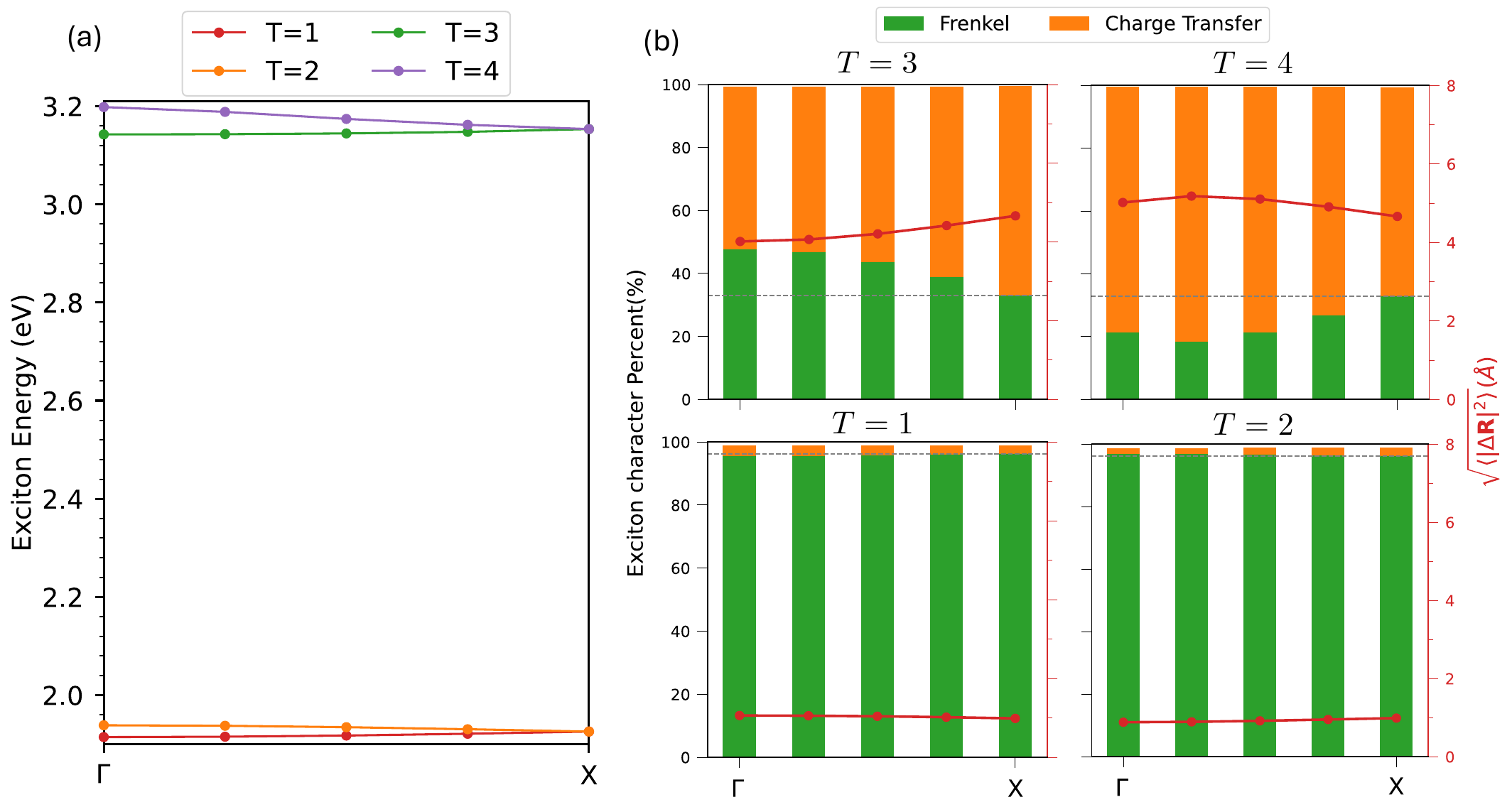}
\caption{\label{fig:anth_Q_T}$\mathbf{Q}$-dependence of triplet exciton spatial character in crystal anthracene.}
\end{figure}

\begin{figure}[H]
\includegraphics[width=1\linewidth]{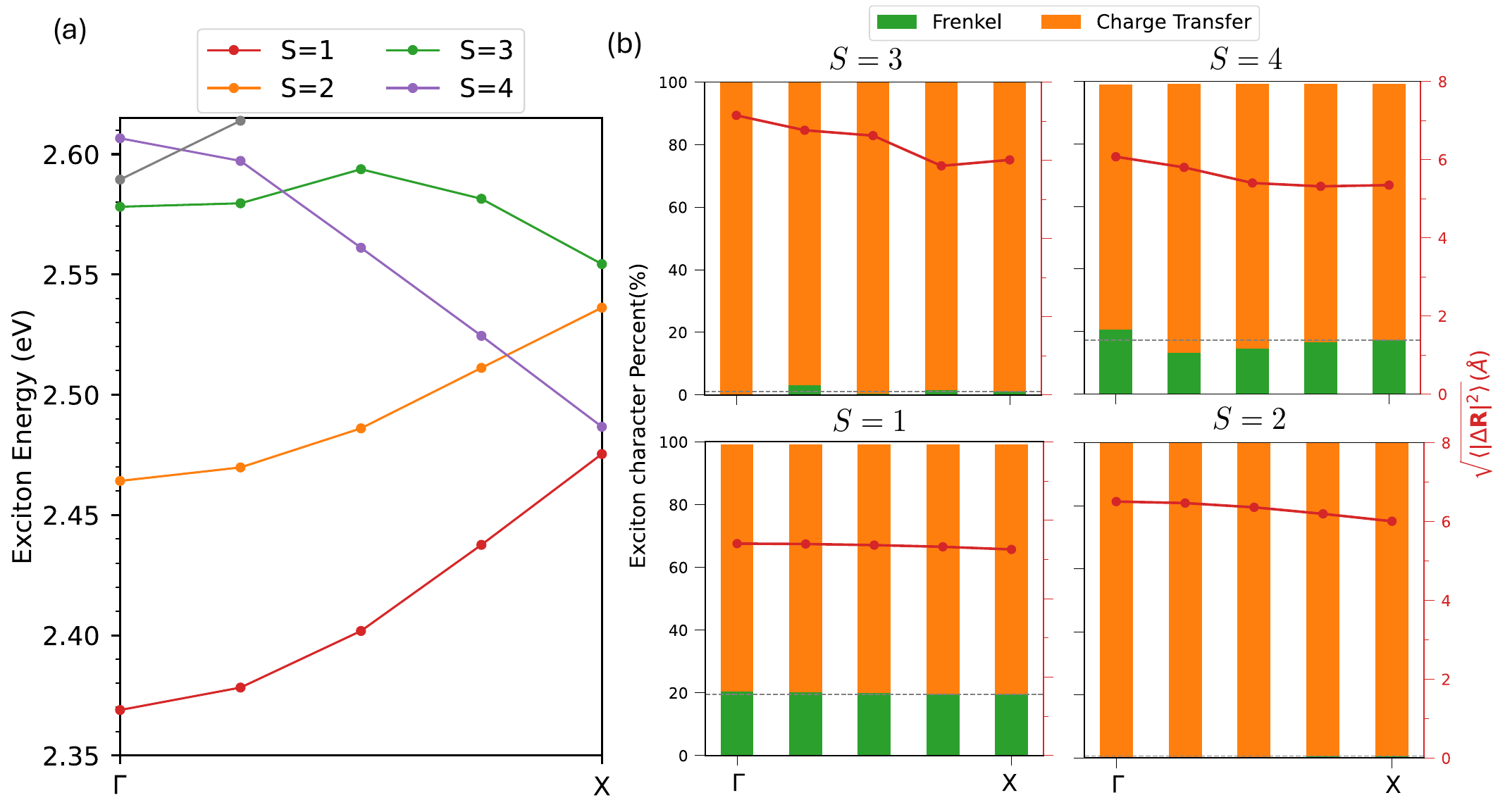}
\caption{\label{fig:tetra_Q_S}$\mathbf{Q}$-dependence of singlet exciton spatial character in crystal tetracene.}
\end{figure}

\begin{figure}[H]
\includegraphics[width=1\linewidth]{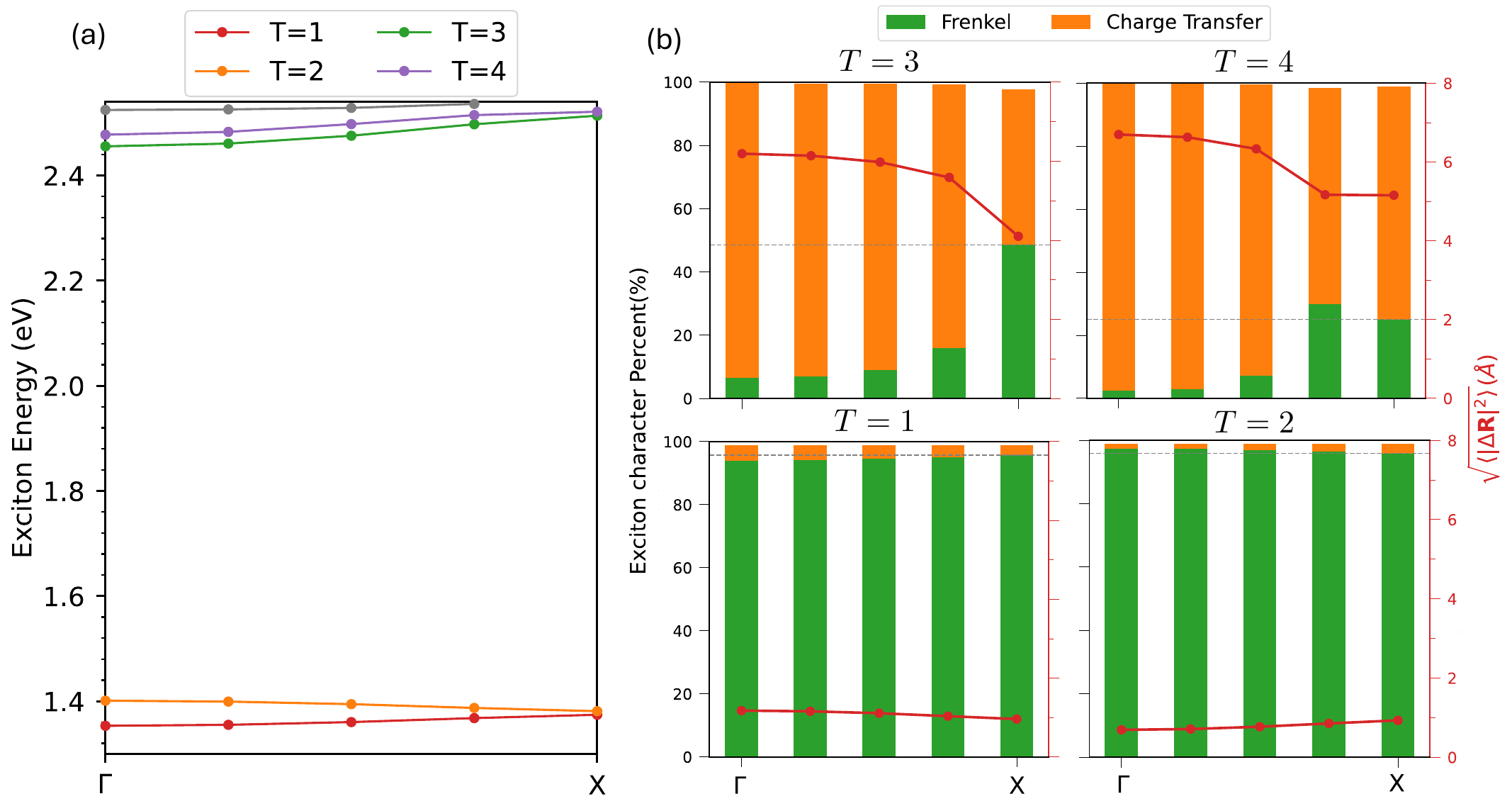}
\caption{\label{fig:tetra_Q_T}$\mathbf{Q}$-dependence of triplet exciton spatial character in crystal tetracene.}
\end{figure}

\begin{figure}[H]
\includegraphics[width=1\linewidth]{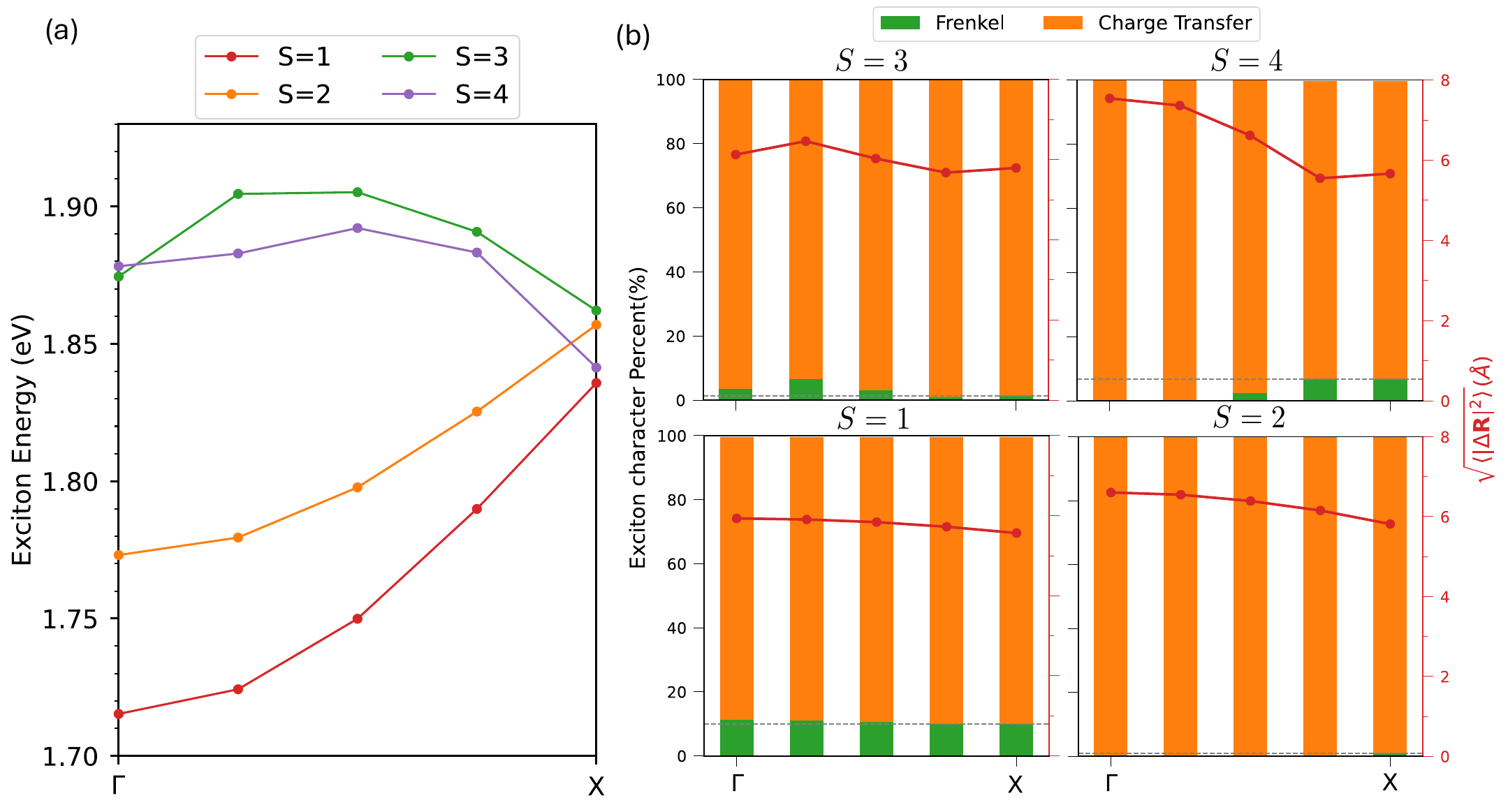}
\caption{\label{fig:penta_Q_S}$\mathbf{Q}$-dependence of singlet exciton spatial character in crystal pentacene.}
\end{figure}

\begin{figure}[H]
\includegraphics[width=1\linewidth]{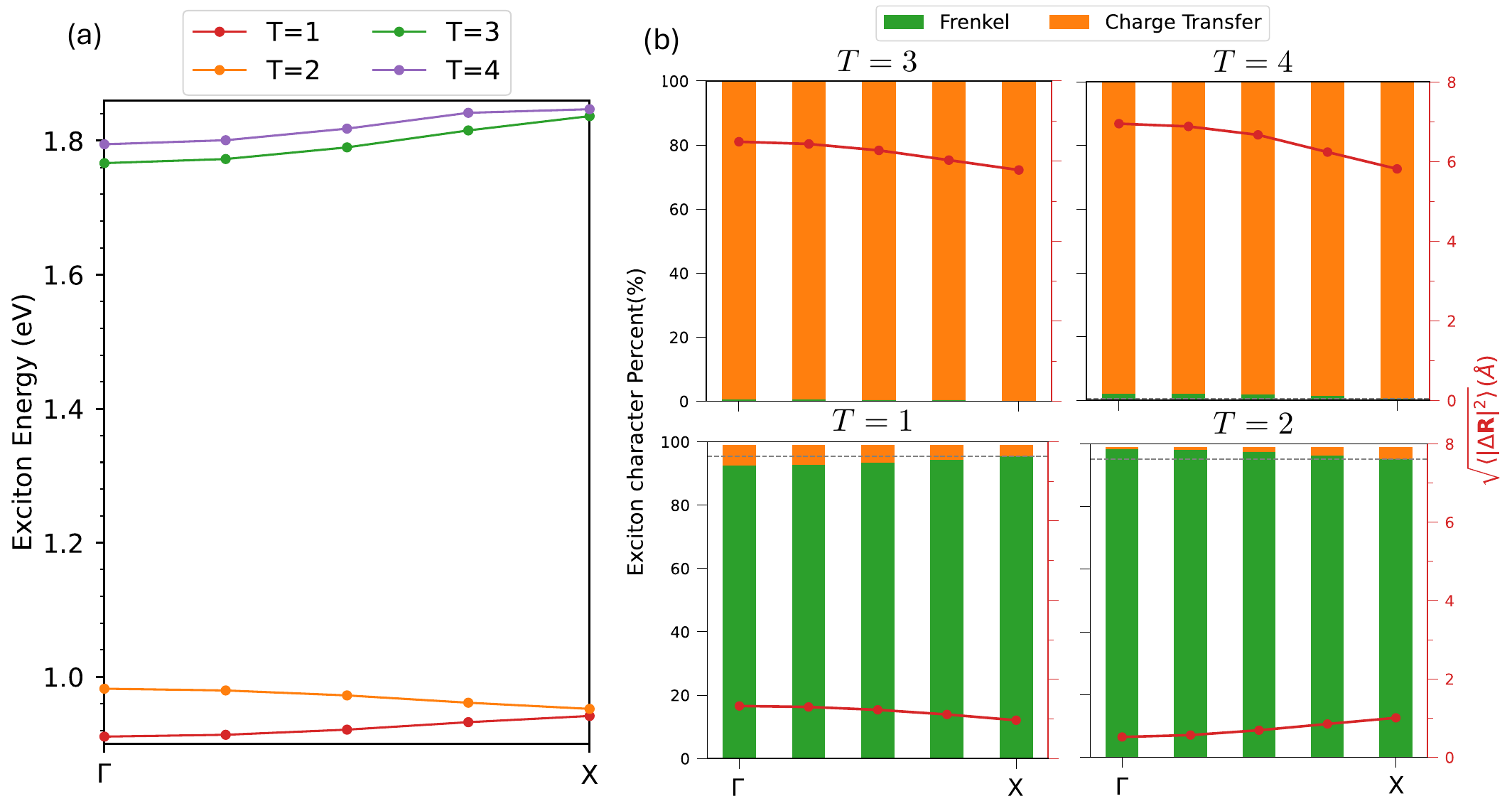}
\caption{\label{fig:penta_Q_T}$\mathbf{Q}$-dependence of triplet exciton spatial character in crystal pentacene.}
\end{figure}

 Fig.~\ref{fig:naph_Q_S} - \ref{fig:penta_Q_T} together with Fig.~\ref{fig:anth_Q} in the main text show: (a) Exciton dispersion for singlet or triplet in acene crystals. (b) The spatial character change for the lowest four excitons along the \textbf{Q}-path.

\section{\label{sec:acene_arrow_plot}WFDX Arrow Plots}

\begin{figure}[H]
\centering
\includegraphics[width=1\linewidth]{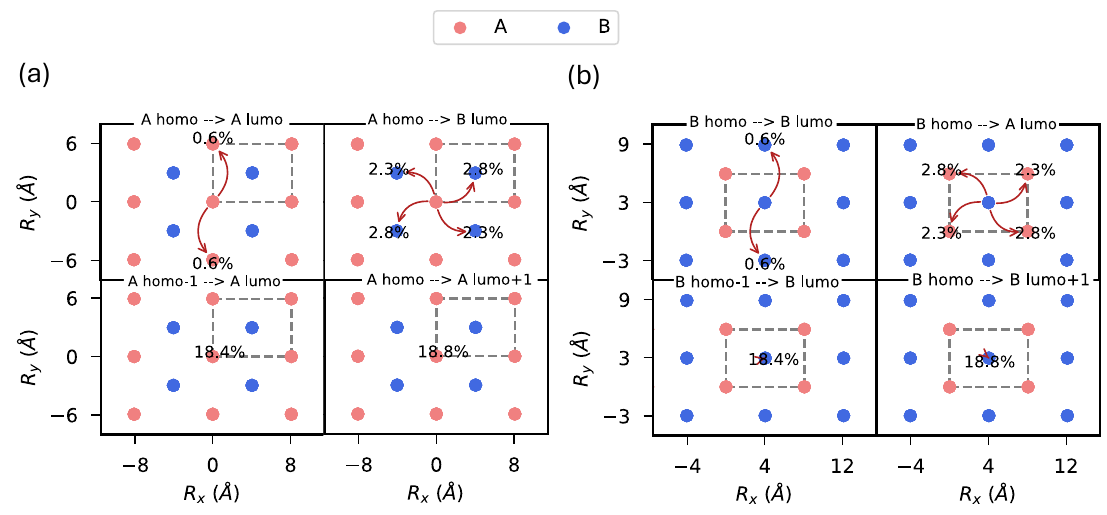}
\caption{\label{fig:naph_S_arrow}Arrow plots for singlet in crystal naphthalene.}
\end{figure}

\begin{figure}[H]
\includegraphics[width=1\linewidth]{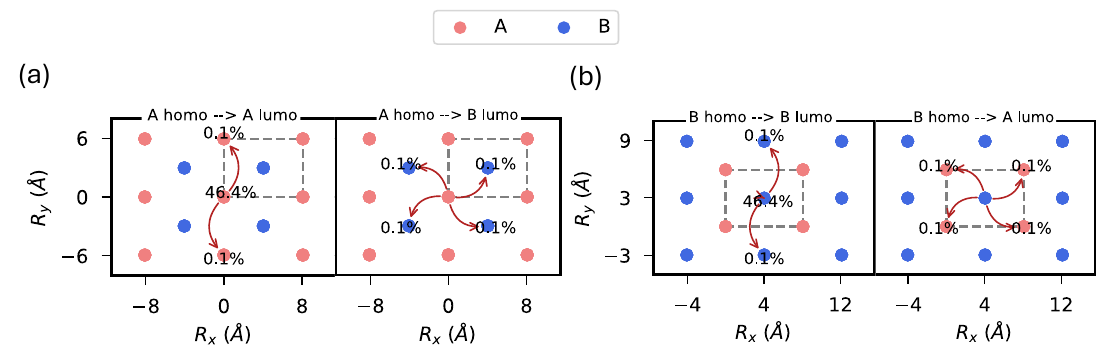}
\caption{\label{fig:naph_T_arrow}Arrow plots for triplet in crystal naphthalene.}
\end{figure}

\begin{figure}[H]
\includegraphics[width=1\linewidth]{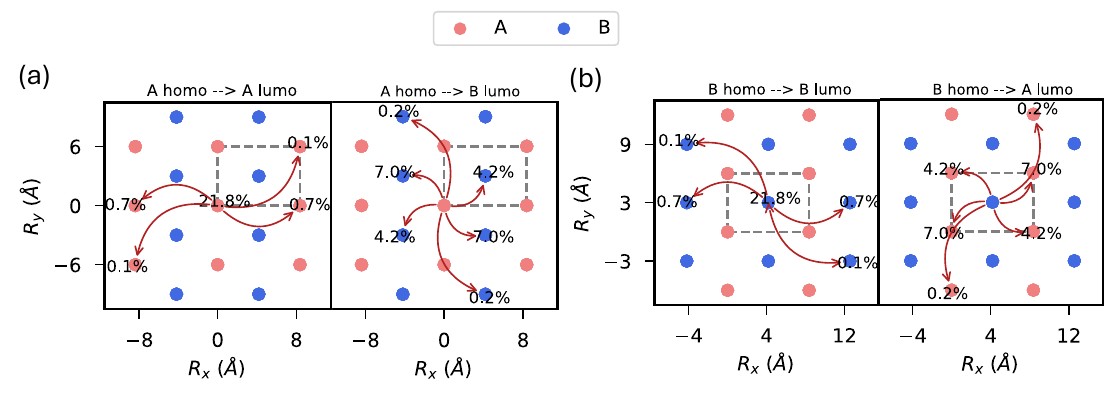}
\caption{\label{fig:anth_S_arrow}Arrow plots for singlet in crystal anthracene.}
\end{figure}

\begin{figure}[H]
\includegraphics[width=1\linewidth]{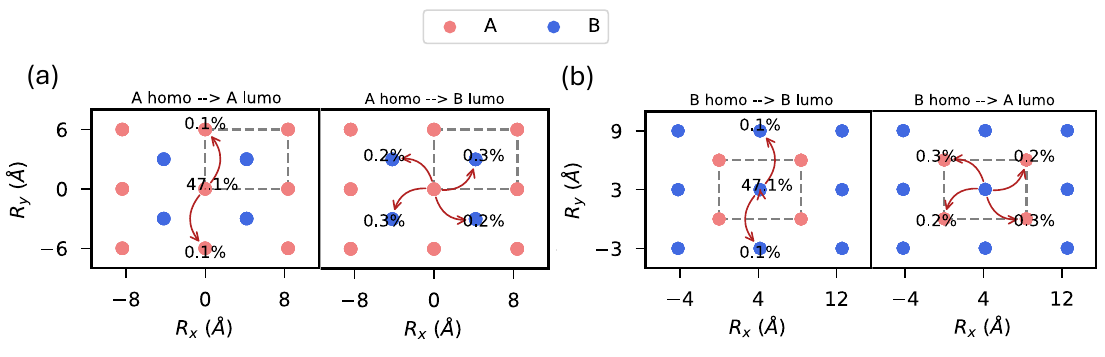}
\caption{\label{fig:anth_T_arrow}Arrow plots for triplet in crystal anthracene.}
\end{figure}

\begin{figure}[H]
\includegraphics[width=1\linewidth]{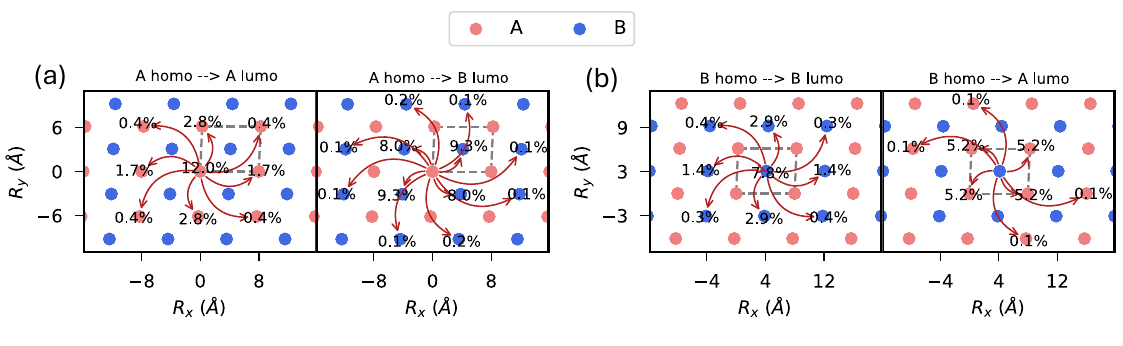}
\caption{\label{fig:tetra_S_arrow}Arrow plots for singlet in crystal tetracene.}
\end{figure}

\begin{figure}[H]
\includegraphics[width=1\linewidth]{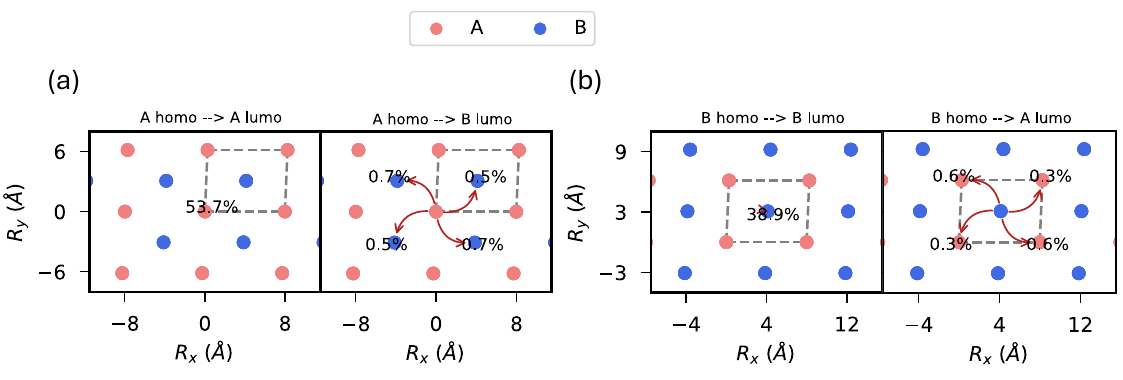}
\caption{\label{fig:tetra_T_arrow}Arrow plots for triplet in crystal tetracene.}
\end{figure}

\begin{figure}[H]
\includegraphics[width=1\linewidth]{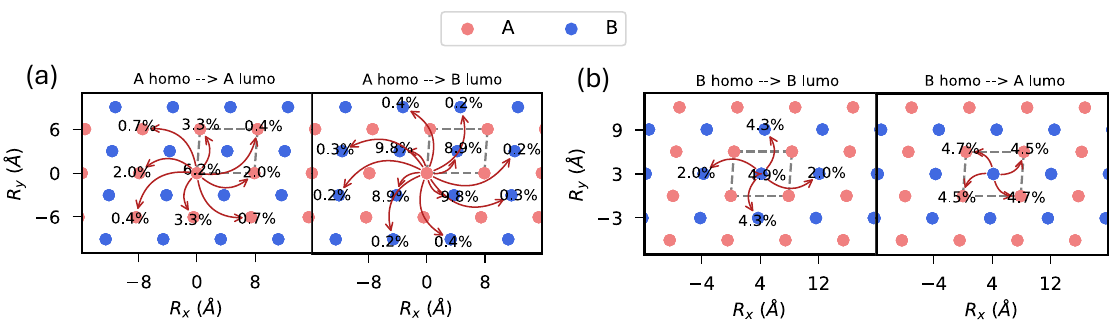}
\caption{\label{fig:penta_S_arrow}Arrow plots for singlet in crystal pentacene.}
\end{figure}

\begin{figure}[H]
\includegraphics[width=1\linewidth]{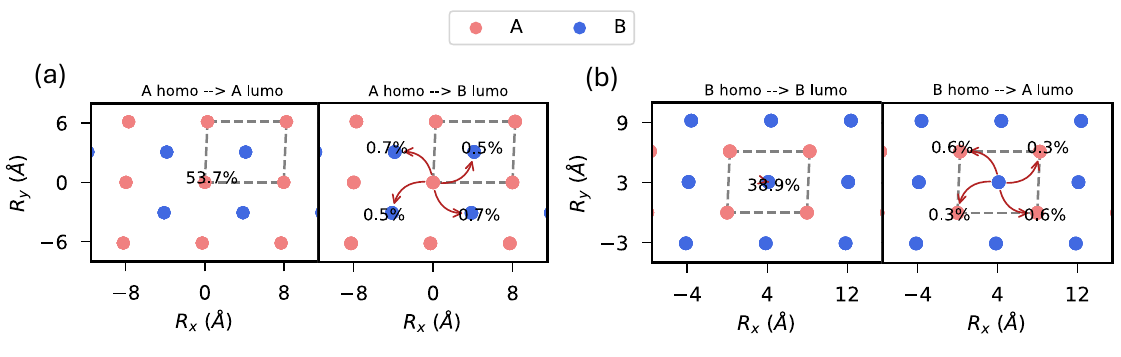}
\caption{\label{fig:penta_T_arrow}Arrow plots for triplet in crystal pentacene.}
\end{figure}

Fig.~\ref{fig:naph_S_arrow} - \ref{fig:penta_T_arrow} present WFDX arrow plots for the lowest excitons at zero $\mathbf{Q}$ momentum ($\mathbf{Q}=0$), both singlet and triplet, across the acene crystals. Summing all annotated probabilities for having specific orbital transitions and electron-hole separation within one figure yields $\approx100\%$, confirming normalization. These plots also provide quick consistency checks: they reveal the growth of real-space exciton delocalization with increasing number of rings and repulsive exchange, and they expose A $\leftrightarrow$ B sublattice relations enforced by the crystals' nonsymmorphic symmetry.

\section{\label{sec:nonsym}Transformation of Exciton Expansion Coefficients under Glide-Mirror Symmetry}
We start from the Wannier basis expansion of the exciton wavefunction,
\begin{equation}
        \Psi_{S \mathbf{Q}}\left(\mathbf{r}_e, \mathbf{r}_h\right)= \frac{1}{\sqrt{N_{\mathbf{k}}}} \sum_{\bar{\mathbf{R}}} e^{i \mathbf{Q} \cdot \bar{\mathbf{R}}} \left[ \sum_{m n \Delta \bar{\mathbf{R}}} A_{m n \Delta \bar{\mathbf{R}}}^{S \mathbf{Q}} w_{m \bar{\mathbf{R}}+\Delta \bar{\mathbf{R}}}\left(\mathbf{r}_e\right) w_{n \bar{\mathbf{R}}}^{\star}\left(\mathbf{r}_h\right)\right],
    \label{eq:xct_mn}
\end{equation}
where $m,n \in \{\mathrm{A,B}\}$ label the two sublattices.  The two nonsymmorphic glide mirror operations are
\begin{equation}
    P_{\perp b} = \{\sigma_{\perp b} \,|\, \tau\}, \, P_{\parallel b }= \{\sigma_{\parallel b} \,|\, \tau \}
    \label{eq:glide_mirror}
\end{equation}
$\sigma_{\perp b}$, $\sigma_{\parallel b}$ denote the mirror lying perpendicular and parallel to the $b$ axis, with matrix form
\begin{equation}
    \overleftrightarrow{\sigma}_{\perp b}=\begin{pmatrix}
1 & \quad0 & \quad0 \quad \\
0 & -1 & \quad0 \quad \\
0 & \quad0 & \quad1 \quad
\end{pmatrix},\label{eq:perp_b}
    \overleftrightarrow{\sigma}_{\parallel b}=\begin{pmatrix}
-1 & \quad0 & \quad0 \\
\quad0 & \quad1 & \quad0 \\
\quad0 & \quad0 & -1 
\end{pmatrix}. 
\end{equation}    
$\tau$ is the translation operation, and it can be written as a vector in fractional coordinate $\boldsymbol{\tau}=(-1/2, -1/2, 0)$. Nonsymmorphic operation acts on real space coordinate give,
\begin{equation}
    \mathbf{r}'  = \{\sigma | \tau\} \mathbf{r} = \overleftrightarrow{\sigma} \cdot \mathbf{r}+\boldsymbol{\tau}.
    \label{eq:r2r'}
\end{equation}

Because the exciton is eigenstates of the Hamiltonian describing the electron and hole interactions in crystals, the exciton wavefunction must inherit all of the system symmetries. The space group of crystal acene structure with nonsymmorphic symmetry is $P2_1/c$, which contains only one-dimensional irreducible representation for all the \textbf{k}-points, so the exciton wavefunction must be transformed under nonsymmorphic operation $P$ as
\begin{equation}
    P\Psi_{S\mathbf{Q}}(\mathbf{r}_e, \mathbf{r}_h)=\pm \,\Psi_{S\mathbf{Q}}(\mathbf{r}_e, \mathbf{r}_h).
    \label{eq:P_wfn_xct}
\end{equation}
Since the exciton wavefunction lives in the configuration space tensored by the electron and hole real space coordinate, $P$ must act on both $\mathbf{r}_e$ and $\mathbf{r}_h$. The basis vector for the two sublattices A and B in the home unit cell are $\boldsymbol{\tau}_{\mathrm{A}} = (0,0,0)$ and $\boldsymbol{\tau}_{\mathrm{B}}=(1/2,1/2,0)$. The A site centered electron Wannier functions transform under $P_{\perp b}$ as:
\begin{equation}
\begin{aligned}
      P_{\perp b} \, w_{\mathrm{A},\bar{\mathbf{R}}+\Delta\bar{\mathbf{R}}}(\mathbf{r}_e)&=P_{\perp b}\, w(\mathbf{r}_e-\bar{\mathbf{R}}-\Delta\bar{\mathbf{R}}-\boldsymbol{\tau}_{\mathrm{A}})=\pm \,w(\mathbf{r}_e'-[\overleftrightarrow{\sigma}_{\perp b}\cdot(\bar{\mathbf{R}}+\Delta\bar{\mathbf{R}}+\boldsymbol{\tau}_{\mathrm{A}})+\boldsymbol{\tau}])\\
    &=  \pm \,w(\mathbf{r}_e'-[\overleftrightarrow{\sigma}_{\perp b}\cdot(\bar{\mathbf{R}}+\Delta\bar{\mathbf{R}})+(-1,-1,0)+\boldsymbol{\tau}_\mathrm{B}])\\
    &= \pm \,w_{\mathrm{B},\overleftrightarrow{\sigma}_{\perp b}\cdot(\bar{\mathbf{R}}+\Delta\bar{\mathbf{R}})+(-1,-1,0)}(\mathbf{r}_e').
    \label{eq:perp_b_e_A}
\end{aligned}
\end{equation}
Here the phase changes of the Wannier function under the space group operation is related to the corresponding molecular orbital symmetry; one can determine this explicitly using the irreducible representation of the molecule's ``local group". An overall $\pm$ sign holds and we keep it in this way for simplicity and universality of orbital type. The phase doesn't matter since at the end we care about the magnitude of the probability amplitude. Similarly, we have
\begin{subequations}
    \begin{align}
        P_{\perp b} \, w_{\mathrm{B},\bar{\mathbf{R}}+\Delta\bar{\mathbf{R}}}(\mathbf{r}_e) &= \pm \,w_{\mathrm{A},\overleftrightarrow{\sigma}_{\perp b}\cdot(\bar{\mathbf{R}}+\Delta\bar{\mathbf{R}})+(0,-1,0)}(\mathbf{r}_e');\label{eq:perp_b_e_B}\\
        P_{\parallel b} \, w_{\mathrm{A},\bar{\mathbf{R}}+\Delta\bar{\mathbf{R}}}(\mathbf{r}_e) &= \pm \,w_{\mathrm{B},\overleftrightarrow{\sigma}_{\parallel b}\cdot(\bar{\mathbf{R}}+\Delta\bar{\mathbf{R}})+(-1,-1,0)}(\mathbf{r}_e'),\label{eq:para_b_e_A}\\
        P_{\parallel b} \, w_{\mathrm{B},\bar{\mathbf{R}}+\Delta\bar{\mathbf{R}}}(\mathbf{r}_e) &= \pm \,w_{\mathrm{A},\overleftrightarrow{\sigma}_{\parallel b}\cdot(\bar{\mathbf{R}}+\Delta\bar{\mathbf{R}})+(-1,0,0)}(\mathbf{r}_e');\label{eq:para_b_e_B}\\
        P_{\perp b} \, w_{\mathrm{A},\bar{\mathbf{R}}}^{\star}(\mathbf{r}_h) &= \pm \,w_{\mathrm{B},\overleftrightarrow{\sigma}_{\perp b}\cdot\bar{\mathbf{R}}+(-1,-1,0)}^{\star}(\mathbf{r}_h'),\label{eq:perp_b_h_A}\\        
        P_{\perp b} \, w_{\mathrm{B},\bar{\mathbf{R}}}^{\star}(\mathbf{r}_h) &= \pm \,w_{\mathrm{A},\overleftrightarrow{\sigma}_{\perp b}\cdot\bar{\mathbf{R}}+(0,-1,0)}^{\star}(\mathbf{r}_h');\label{eq:perp_b_h_B}\\
        P_{\parallel b} \, w_{\mathrm{A},\bar{\mathbf{R}}}^{\star}(\mathbf{r}_h) &= \pm \,w_{\mathrm{B},\overleftrightarrow{\sigma}_{\parallel b}\cdot\bar{\mathbf{R}}+(-1,-1,0)}^{\star}(\mathbf{r}_h'),\label{eq:para_b_h_A}\\
        P_{\parallel b} \, w_{\mathrm{B},\bar{\mathbf{R}}}^{\star}(\mathbf{r}_h) &= \pm \,w_{\mathrm{A},\overleftrightarrow{\sigma}_{\parallel b}\cdot\bar{\mathbf{R}}+(-1,0,0)}^{\star}(\mathbf{r}_h').\label{eq:para_b_h_B}
    \end{align}
    \label{eq:p_b_eh}
\end{subequations}
The space operation $P$ acts on electron-hole Wannier center difference $\Delta \mathbf{R}$ as
\begin{equation}
        P\,\Delta \mathbf{R}=\overleftrightarrow{\sigma} \cdot \Delta\bar{\mathbf{R}}+(\overleftrightarrow{\sigma} \cdot \boldsymbol{\tau}_m+\boldsymbol{\tau})-(\overleftrightarrow{\sigma} \cdot \boldsymbol{\tau}_n+\boldsymbol{\tau})= \overleftrightarrow{\sigma} \cdot \Delta \mathbf{R}. 
        \label{eq:P_DeltaR}
\end{equation}
Checking this relation using the result in Eq.~\ref{eq:perp_b_e_A} and~\ref{eq:p_b_eh}: for $m=\mathrm{A}, n=\mathrm{A}$, $P_{\perp b}$ acts on $\Delta \mathbf{R}$ can be evaluated using Eq.~\ref{eq:perp_b_e_A} and~\ref{eq:perp_b_h_A}
\begin{equation}
\begin{aligned}
    P_{\perp b}: \Delta \mathbf{R}&= (\bar{\mathbf{R}}+\Delta \bar{\mathbf{R}}+\boldsymbol{\tau}_{\mathrm{A}})-(\bar{\mathbf{R}}+\boldsymbol{\tau}_{\mathrm{A}})=\Delta \bar{\mathbf{R}} \\
    \mapsto \Delta \mathbf{R}' &= [\overleftrightarrow{\sigma}_{\perp b} \cdot (\bar{\mathbf{R}}+\Delta \bar{\mathbf{R}})+(-1,-1,0)+\boldsymbol{\tau}_{\mathrm{B}}]-[\overleftrightarrow{\sigma}_{\perp b} \cdot\bar{\mathbf{R}}+(-1,-1,0)+\boldsymbol{\tau}_{\mathrm{B}}]\\
    &= \overleftrightarrow{\sigma}_{\perp b} \cdot \Delta \bar{\mathbf{R}} =  \overleftrightarrow{\sigma}_{\perp b} \cdot \Delta \mathbf{R}.
    \label{eq:P_DeltaR_AA}
\end{aligned}
\end{equation}
For $m=\mathrm{A}, n=\mathrm{B}$, $P_{\perp b}$ acts on $\Delta \mathbf{R}$ can be evaluated using Eq.~\ref{eq:perp_b_e_A} and~\ref{eq:perp_b_h_B}
\begin{equation}
\begin{aligned}
    P_{\perp b}: \Delta \mathbf{R}&= (\bar{\mathbf{R}}+\Delta \bar{\mathbf{R}}+\boldsymbol{\tau}_{\mathrm{A}})-(\bar{\mathbf{R}}+\boldsymbol{\tau}_{\mathrm{B}})=\Delta \bar{\mathbf{R}}+\boldsymbol{\tau}_{\mathrm{A}}-\boldsymbol{\tau}_{\mathrm{B}}=\Delta \bar{\mathbf{R}}+(-1/2,-1/2,0) \\
    \mapsto \Delta \mathbf{R}' &= [\overleftrightarrow{\sigma}_{\perp b} \cdot (\bar{\mathbf{R}}+\Delta \bar{\mathbf{R}})+(-1,-1,0)+\boldsymbol{\tau}_{\mathrm{B}}]-[\overleftrightarrow{\sigma}_{\perp b} \cdot\bar{\mathbf{R}}+(0,-1,0)+\boldsymbol{\tau}_{\mathrm{A}}]\\
    &= \overleftrightarrow{\sigma}_{\perp b} \cdot \Delta \bar{\mathbf{R}}+(-1/2,1/2,0) =  \overleftrightarrow{\sigma}_{\perp b} \cdot (\Delta \bar{\mathbf{R}}+\boldsymbol{\tau}_{\mathrm{A}}-\boldsymbol{\tau}_{\mathrm{B}})=\overleftrightarrow{\sigma}_{\perp b} \cdot \Delta \mathbf{R}.
    \label{eq:P_DeltaR_AA}
\end{aligned}
\end{equation}
Similarly for others. 

Define $\Delta \mathbf{R}'=\overleftrightarrow{\sigma}\cdot \Delta \mathbf{R}$, $\bar{m} = P \,m = P \,\mathrm{A(B)} = \mathrm{B(A)}$, $\bar{n} = P \,n = P \,\mathrm{A(B)} = \mathrm{B(A)}$; new symbol $\mathbf{R}_n$ and $\mathbf{R}_{\bar{n}}=P \, \mathbf{R}_n$ indicating the Wannier center coordinate of the hole on sublattice $n$ and $\bar{n}$. The subscript $n, m, \bar{n}, \bar{m}$ is still kept to emphasize on the sublattice where the Wannier center lies. We can get a general form for Wannier functions transformed under $P$ as
\begin{subequations}
\begin{align}
    P &\, w_{n} (\mathbf{r}-\bar{\mathbf{R}}_n)=\pm \,w_{\bar{n}}(\mathbf{r}'-\bar{\mathbf{R}}_{\bar{n}}),
    \label{eq:general_P_WF_h}\\
    P &\, w_{m} (\mathbf{r}-\bar{\mathbf{R}}_n-\Delta \mathbf{R})=\pm \,w_{\bar{m}}(\mathbf{r}'-\bar{\mathbf{R}}_{\bar{n}}-\Delta \mathbf{R}').
    \label{eq:general_P_WF_e}        
\end{align}
\end{subequations}
Applying $P$ to the exciton wavefunction yields:
\begin{equation}
    \begin{aligned}
        P \Psi_{S\mathbf{Q}}(\mathbf{r}_e,\mathbf{r}_h)&=P \frac{1}{\sqrt{N_{\mathbf{k}}}} \sum_{\bar{\mathbf{R}}} e^{i \mathbf{Q} \cdot \bar{\mathbf{R}}} \left[ \sum_{m n \Delta \bar{\mathbf{R}}} A_{m n \Delta \bar{\mathbf{R}}}^{S \mathbf{Q}} w_{m \bar{\mathbf{R}}+\Delta \bar{\mathbf{R}}}\left(\mathbf{r}_e\right) w_{n \bar{\mathbf{R}}}^{\star}\left(\mathbf{r}_h\right)\right]\\
        &=P \frac{1}{\sqrt{N_{\mathbf{k}}}} \sum_{\bar{\mathbf{R}}} e^{i \mathbf{Q} \cdot \bar{\mathbf{R}}} \left[ \sum_{m n \Delta \mathbf{R}} A_{m n \Delta \mathbf{R}}^{S \mathbf{Q}} w_{m }\left(\mathbf{r}_e-\bar{\mathbf{R}}_n-\Delta \mathbf{R}\right) w_{n }^{\star}\left(\mathbf{r}_h-\bar{\mathbf{R}}_n\right)\right]\\
        &=\pm \,\frac{1}{\sqrt{N_{\mathbf{k}}}} \sum_{\bar{\mathbf{R}}} e^{i \mathbf{Q} \cdot \bar{\mathbf{R}}} \left[ \sum_{\bar{m} \bar{n} \Delta \mathbf{R}'} A_{m n \Delta \mathbf{R}}^{S \mathbf{Q}} w_{\bar{m} }\left(\mathbf{r}_e'-\bar{\mathbf{R}}_{\bar{n}}-\Delta \mathbf{R}'\right) w_{\bar{n} }^{\star}\left(\mathbf{r}_h'-\bar{\mathbf{R}}_{\bar{n}}\right)\right]\\
        &=\pm \,\frac{1}{\sqrt{N_{\mathbf{k}}}} \sum_{\bar{\mathbf{R}}} e^{i \mathbf{Q} \cdot \bar{\mathbf{R}}} \left[ \sum_{m n \Delta \mathbf{R}} A_{\bar{m} \bar{n} \Delta \mathbf{R}'}^{S \mathbf{Q}} w_{m}\left(\mathbf{r}_e-\bar{\mathbf{R}}_n-\Delta \mathbf{R}\right) w_{n}^{\star}\left(\mathbf{r}_h-\bar{\mathbf{R}}_n\right)\right]\\
        &=\pm \,\frac{1}{\sqrt{N_{\mathbf{k}}}} \sum_{\bar{\mathbf{R}}} e^{i \mathbf{Q} \cdot \bar{\mathbf{R}}} \left[ \sum_{m n \Delta \mathbf{R}} A_{m n \Delta \mathbf{R}}^{S \mathbf{Q}} w_{m }\left(\mathbf{r}_e-\bar{\mathbf{R}}_n-\Delta \mathbf{R}\right) w_{n }^{\star}\left(\mathbf{r}_h-\bar{\mathbf{R}}_n\right)\right],
    \end{aligned}
    \label{eq:coeff_=}
\end{equation}
where line 3 to line 4 results from relabeling $\bar{m}$ as $m$, $\bar{n}$ as $n$, $\bar{\mathbf{R}}_{\bar{n}}$ as $\bar{\mathbf{R}}_n$ and $\Delta \mathbf{R}'$ as $\Delta \mathbf{R}$. The final line is a reform of Eq.~\ref{eq:P_wfn_xct}. We therefore proved that
\begin{equation}
    \left|A_{m n \Delta \mathbf{R}}^{S \mathbf{Q}}\right|^2=\left|A_{\bar{m} \bar{n} ,\, \overleftrightarrow{\sigma}\cdot\Delta \mathbf{R}}^{S \mathbf{Q}}\right|^2.
    \label{eq:eqaul_coeff}
\end{equation}

% \nocite{*}
\twocolumngrid
\bibliography{main}% Produces the bibliography via BibTeX.
\end{document}